\documentclass[12pt]{article}

\usepackage{amsmath}		
\usepackage{amssymb}

\usepackage{microtype}		
\usepackage{slashed}		
\usepackage{cite}			
\usepackage{hyperref}
\hypersetup{
	bookmarksnumbered,
	unicode,			
	colorlinks,			
	citecolor=[rgb]{.9,0,.5},	
	urlcolor=[rgb]{0,0,1},	
	linkcolor=[rgb]{0,.7,0}	
	}

\usepackage{empheq}
\usepackage[most]{tcolorbox}					
\definecolor{realcyan}{rgb}{0,1,1}
\definecolor{realyellow}{rgb}{1,1,0}
\newtcbox{\mymath}[1][]{%
    nobeforeafter, math upper, tcbox raise base,
    enhanced, colframe=realcyan!50!white,
    colback=realyellow!5!white, boxrule=1pt, drop lifted shadow, sharp corners
    #1}

\parskip=\medskipamount				

\numberwithin{equation}{section}

\mathcode`\*="702A                  	

\catcode`@=11
\let\over\@@over
\catcode`@=12

\def\on#1#2{{\buildrel{{\mkern2.5mu\raise-.1em\hbox{$\scriptstyle#1$}\mkern-2.5mu}}\over{#2}}}
\def\ron#1#2{{\buildrel{{\raise-.1em\hbox{$\scriptstyle#1$}}}\over{#2}}}	
\def\ddt#1{\on{\hbox{\bf .\kern-1pt.}}#1}							

\providecommand{\smalltriangleright}{\vartriangleright}
\def\dd{\hbox{\large$\smalltriangleright$}}

\def\dig#1{\setbox0=\hbox{$#1M$}
	\hskip.06\wd0 \vrule width.07\wd0 height.63\wd0 depth.01\wd0
	\vrule width.37\wd0 height.63\wd0 depth-.56\wd0 \hskip-.4\wd0
	\vrule width.25\wd0 height.35\wd0 depth-.28\wd0
	\vrule width.07\wd0 height.35\wd0 depth-.17\wd0 \hskip.14\wd0}
\def\digamma{{\mathpalette\dig{}}}

\def\rm{\mathrm}	

\def\l{{\ell}}\def\={{\;=\;}}\def\+{{\;+\;}}

\def\don{\on}

\newcommand{\bea}{\begin{eqnarray}}
\newcommand{\eea}{\end{eqnarray}}
\newcommand{\bref}[1]{(\ref{#1})}
\newcommand{\nn}{\nonumber}
\usepackage{multirow}

\begin{document}
	
\hfill KEK-TH-2337
\vskip-.1in
\hfill YITP-SB-2021-11

{\center
	{\color{cyan}
		\fontsize{28pt}{34pt}\bf\sffamily
		Perturbative F-theory 10-brane and M-theory 5-brane\\[.5in]
	}

\href{mailto:mhatsuda@juntendo.ac.jp}{Machiko Hatsuda} \\[.1in]
{\it
Department of Radiological Technology, Faculty of Health Science, Juntendo University\\
Yushima, Bunkyou-ku, Tokyo 113-0034, Japan\\
KEK Theory Center, High Energy Accelerator Research Organization\\
Tsukuba, Ibaraki 305-0801, Japan \\[.1in]
}
and \\[.1in]
\href{mailto:siegel@insti.physics.sunysb.edu}{Warren Siegel} \\[.1in]
{\it
\href{http://insti.physics.sunysb.edu/~siegel/plan.html}{C. N. Yang Institute for Theoretical Physics}\\
State University of New York, Stony Brook, NY 11794-3840}\\[.3in]

{\color[rgb]{0,.7,1}\today}\\[.5in]

}

{\abstract
The exceptional symmetry is realized perturbatively in F-theory which is the manifest U-duality theory.
The SO(5,5) U-duality symmetry acts on both the 16 spacetime coordinates  and the 10 worldvolume coordinates.
Closure of the Virasoro algebra requires the Gauss law constraints on the worldvolume.
This set  of current algebras describes a F-theory 10-brane.
The SO(5,5) duality symmetry is enlarged to the SO(6,6) symmetry in the Lagrangian formulation.
We propose actions of the F-theory 10-brane with SO(5,5) and SO(6,6) symmetries.
The gauge fields of the 
latter action are coset elements of SO(6,6)/SO(6;$\mathbb{C}$) which include
both the  SO(5,5)/SO(5;$\mathbb{C}$) spacetime backgrounds and the worldvolume backgrounds.
The SO(5,5) current algebra obtained from the Pasti-Sorokin-Tonin M5-brane Lagrangian
leads to the theory behind M-theory, namely F-theory.
We  also propose an action of the perturbative M-theory 5-brane
obtained by sectioning the worldvolume of the F-theory 10-brane. 
}

\vfill

\thispagestyle{empty}
\newpage

{
\tableofcontents
}

\newpage
\section{Introduction}
Duality is a cornucopia of string theory creating its unique properties. Superstring theory is considered to be a candidate for a unified theory of all forces, 
and five types of superstring theories have been shown to exist.
Five superstring theories together with M-theory 
form pairs related by T duality or S duality.
Then why are superstring theories related  in a way of chain of dualities?
T duality is the equivalence under the interchange of  
$R\leftrightarrow \alpha'/R$ for the radius of the compactified space $R$,
while S duality is the equivalence under the  interchange of  
$g\leftrightarrow 1/g$ for the string coupling $g$. So they relate paired theories. 
These T duality and S duality are encompassed by U duality. 
Therefore we examine  whether a theory with manifest U-duality exists which describes different superstring theories by its different sections. We call such a theory with manifest U-duality  ``F-theory".

The theory with manifest T-duality was presented in \cite{Siegel:1993bj,Siegel:1993th,Siegel:1993xq}
which was named ``T-theory" later.
``T-theory" was defined by the O($D,D$) current algebra
which generates gauge symmetries of background gauge fields.
It contains winding modes even in an uncompactified space,
and  the $D$-dimensional space includes time direction
to describe dynamical gravity.
All string modes including massive modes are described in  
the T-theory which has been developed  \cite{Siegel:1993bj,Siegel:1993th,Siegel:1993xq,Polacek:2013nla,Hatsuda:2014aza,Hatsuda:2014qqa,Polacek:2014cva,Hatsuda:2015cia,Hatsuda:2017tfa,Hatsuda:2018tcx,{Hatsuda:2019xiz}}.
Section conditions eliminate winding modes while the string field theory condition
$L_0=\bar{L}_0$ mixes massive winding modes with massive oscillator states \cite{Hull:2009mi}.

Duality relates nonperturbative states of classical string theory.
When duality symmetry is represented linearly in some classical theory,
nonperturbative states are described perturbatively. 
The  O($D,D$) T-duality symmetry is represented linearly in 
the Double Field Theory (DFT) \cite{Siegel:1993bj,Siegel:1993th,Siegel:1993xq,Hull:2009mi,Hull:2009zb,Zwiebach:2011rg,Berman:2013eva,Aldazabal:2013sca,Hohm:2013bwa,Park:2016sbw} and the
generalized geometry  
gives a mathematical framework of geometry induced by T-duality
\cite{{Hitchin:2003cxu},{Hitchin:2005in},{Gualtieri:2003dx}}.
They are theories of massless modes of string theory.

T-duality and S-duality  are unified into U-duality 
by the exceptional symmetry group 
which involves non-perturbative branes   \cite{Hull:1994ys}.
M-theory was conjectured as a theory  to unify superstring theories 
through dualities whose low energy effective theory is the 11-dimensional supergravity theory \cite{Witten:1995ex}.
F-theory was firstly proposed by Vafa 
\cite{Vafa:1996xn} to understand the IIB theory in the string duality web where similar ideas are also referred \cite{{Blencowe:1988sk},{Hull:1995xh}}.
The  O($D,D$) T-duality symmetry is extended to  the  exceptional symmetry groups U-duality symmetry in
generalized geometry for M-theory 
\cite{Hull:2007zu,PiresPacheco:2008qik}
and   Exceptional Field Theory (EFT) \cite{{Berman:2010is},{Berman:2011pe},Coimbra:2011ky,Berman:2012vc,Hohm:2013vpa,Hohm:2013uia,Hohm:2014fxa,Godazgar:2014nqa,Musaev:2015ces,Abzalov:2015ega}.  
Current algebras for branes were calculated to present 
 generalized brackets
and to derive background gauge symmetries
\cite{Hatsuda:2012uk,{Hatsuda:2012vm},{Hatsuda:2013dya}} corresponding to 
the U-duality covariant formulation 
of  the 11-dimensional supergravity  \cite{{Berman:2010is},{Berman:2011pe}}.

Recently ``F-theory" as a theory with manifest U-duality has been realized \cite{Polacek:2014cva,Linch:2015fca,Linch:2015lwa,Linch:2015fya,Linch:2015qva,Ju:2016hla,Linch:2016ipx,{Siegel:2016dek},{Siegel:2019wrr},{Siegel:2018puf},{Siegel:2020qef}} as a generalization of T-theory.
 F-theory is defined by the exceptional group current algebras
 on branes. 
The exceptional group acts both the spacetime coordinates
and the worldvolume coordinates.
Since the exceptional group includes both the O($D,D$) T-duality symmetry
and the SL(2;$\mathbb{R}$) S-duality symmetry,
F-theory reduces both the IIB theory and M-theory directly
by sectioning or dimensional reduction.
This solves the puzzle of the IIB theory in the duality web  
as shown in the duality diamond \bref{FTMSWeb} including F-theory.

In this paper we focus on the E$_5$=SO(5,5) F-theory.
The SO(5,5) current algebra is realized by a 10-brane.
The spacetime coordinate is the 16-dimensional  spinor representation of SO(5,5)
 while  
the worldvolume coordinate is 10-dimensional vector representation of it. 
16 is decomposed into $5+10+1$ under GL(5) symmetry
where 5, 10, 1 correspond to the 5-dimensional momentum,
the M2-brane winding mode, the M5-brane winding mode respectively.
This is a generalization of doubling the spacetime coordinate
for linear realization of the O($D,D$) T-duality symmetry as $D$ momenta plus $D$ winding modes.
We
propose two different ways of writing the F-theory 10-brane actions:
1. the Hamiltonian form action and 2. Lagrangian form action.
\begin{enumerate}
	\item{ 
The Hamiltonian form action is based on the SO(5,5) ``G-symmetry" current algebra. 
The Lagrangian is written in terms of the 
selfdual and anti-selfdual field strengths,  $\don\circ{F}_{\rm{SD}}{}^\mu$ and $\don\circ{F}_{\overline{\rm{SD}}}{}^\mu$ with $\mu=1,\cdots,16$.
The background gauge  fields $G_{\mu\nu}$ are coset elements of G/H
where  H is a subgroup of G. 
The worldvolume index is $\underline{m}=1,\cdots,10$ and $\gamma_{\underline{m}\mu\nu}$ is the 10-dimensional gamma matrix.
We propose the SO(5,5) symmetric F-theory 10-brane action in curved backgrounds in
\bref{FSDaction} as
\bea
I&=&\int d\tau d^{10}\sigma~L\nn\\
L&=&\frac{1}{g}
\don\circ{F}_{\rm{SD}}{}^\mu G_{\mu\nu}
\don\circ{F}_{\overline{\rm{SD}}}{}^\nu
-\hat{\lambda}
\don\circ{F}_{\overline{\rm{SD}}}{}^\mu G_{\mu\nu}
\don\circ{F}_{\overline{\rm{SD}}}{}^\nu
-\lambda^{\underline{m}}
\don\circ{F}_{\overline{\rm{SD}}}{}^\mu{\gamma}_{\underline{m}\mu\nu}
\don\circ{F}_{\overline{\rm{SD}}}{}^\nu
\eea
where $g$, $\hat{\lambda}$ and  $\lambda^{\underline{m}}$
are Lagrange multipliers.}
\item{
In the Lagrangian formalism the SO(5,5) U-duality symmetry is enlarged to SO(6,6) ``F-symmetry".
The field strength is a 12-dimensional Weyl spinor  $\don\circ{F}_{\underline{\mu}}$ with $\underline{\mu}=1,\cdots,32$. 
The background gauge fields $G^{\underline{\mu\nu}}$ are coset elements of F/L where L is a subgroup of F.
The worldvolume index is $\underline{\hat{m}}=1,\cdots,12$
and $\Sigma^{\underline{\hat{m}}}$ is the 12-dimensional gamma matrix with $\Sigma^{\underline{\hat{m}\hat{n}}}=\frac{1}{2} \Sigma^{[\underline{\hat{m}}}\Sigma^{\underline{\hat{n}}]}
$. 
We propose the SO(6,6) symmetric F-theory 10-brane action in curved backgrounds in
\bref{FSO66}
as
\bea
I&=&\int d^{12}\sigma ~L\nn\\
L&=&e\don\circ{F}{}_{\underline{\mu}}
G^{\underline{\mu\nu}} \don\circ{F}{}_{\underline{\nu}}
-\frac{1}{2}\lambda_{\underline{\hat{m}\hat{n}}}\don\circ{F}{}_{\underline{\mu}}
(C\Sigma^{\underline{\hat{m}\hat{n}}})^{\underline{\mu\nu}} \don\circ{F}{}_{\underline{\nu}}\label{FSO661}
\eea
where $e$ and ${\lambda}_{\underline{\hat{m}\hat{n}}}$ are Lagrange multipliers.
}
\end{enumerate}

The organization of the paper is as follows.
In the next section we present complete sets of the SO(5,5) current algebras in both the SO(5,5) spinor representation in subsection 2.1 
and the GL(5) tensor representation in subsection 2.2.
The former reveals the structure of the current algebra such as the bosonic $\kappa$-symmetry, while the later gives direct coupling to the 11-dimensional supergravity background.
One of the author presented the SO(5,5) current algebra of the M5-brane \cite{Hatsuda:2013dya}  obtained 
from the Pasti-Sorokin-Tonin(PST) M5-brane Lagrangian \cite{Pasti:1997gx}.
This SO(5,5)  current algebra is recognized as the F-theory 10-brane current algebra 
by doubling the worldvolume coordinate as shown in \bref{SO55GL5tensor}, \bref{SO55M} and \bref{Mst}.
In subsection 3.1 we begin by reviewing the double zweibein method to obtain the 
worldsheet covariant action\cite{{Hatsuda:2018tcx},{Hatsuda:2019xiz}} as a method 
to overcome the problem of the chiral action \cite{Siegel:1983es}.
By applying this method to F-theory we obtain the SO(5,5) F-theory 10-brane action in 
the Hamiltonian formalism in subsection 3.2.
We extend it to the SO(6,6) F-theory 10-brane action in subsection 3.3. 
In the SO(6,6) Lagrangian the worldvolume vielbein merges with the spacetime vielbein. 
In subsection 3.4 we present F-theory 10-brane actions in terms of GL(6) and GL(5) tensors
 to couple the supergravity background. 
In section 4 we present an action for a 5-brane
obtained from the F-theory 10-brane action. 
5 worldvolume dimensions are reduced by solving the 
worldvolume section constraint  ${\cal V}=\partial^m\bar{\partial}_m=0$.
The obtained action for a M-theory 5-brane is sum of the free kinetic term and bilinears of the selfduality constraint.

\par \vskip 6mm

\section{Introduction to F-theory}
We begin by an introduction to ``theories" with manifest dualities such as T-theory and F-theory together with 
``S-theory" and ``M-theory".
S-theory is a string theory compactified to $D$-dimensions
and 
M-theory  is a brane theory  compactified to $(D+1)$-dimensions. 
``Theories"  are 
defined by current algebras with G-symmetry in the Hamiltonian formalism.
The background gauge fields are parameters of cosets G/H
which are generalization of the GL($D$)/SO($D-1$,1) for the vielbein gauge field of the Einstein gravity.
All bosonic component fields are representation of G,
while fermionic fields are representation of H.
A new duality web given in the diamond diagram in \bref{FTMSWeb}
\cite{Linch:2015qva}.
\bea
&\begin{array}{c}
	\begin{array}{c}
		\rm{F}\mathchar`-\rm{theory}\\		\rm{E}_{{D}+1({D}+1)}/\rm{H}_{{D}}\\
		\rm{bispinor}
	\end{array}
	\\
	\swarrow\quad\quad\quad\quad\quad\quad\quad\quad
	\searrow\\
	\begin{array}{c}
		\rm{M}\mathchar`-\rm{theory}\\
		\rm{GL}(D+1)/\rm{SO}(D,1)\\
		{D}+1
	\end{array}
	\quad\quad\quad
	\begin{array}{c}
		\rm{T}\mathchar`-\rm{theory}\\
		\rm{O}(D,D)/\rm{SO}(D-1,1)^2\\
		2{D}
	\end{array}\\
	\searrow\quad\quad\quad\quad\quad\quad\quad\quad\swarrow\\
	\begin{array}{c}
		\rm{S}\mathchar`-\rm{theory}\\
		\rm{GL}(D)/\rm{SO}(D-1,1)\\
		{D}
	\end{array}
\end{array}\label{FTMSWeb}&\nn\\
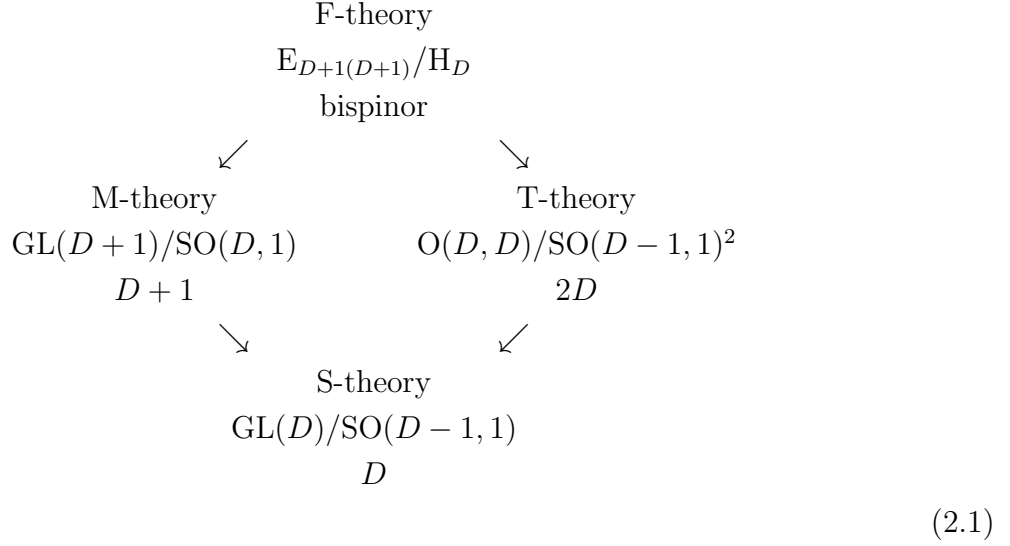
\\&\rm{Figure: ~Diamond~diagram}&\nn
\eea

Theories are also defined by worldvolume actions.
The Hamiltonian form action is obtained from the G-symmetry current algebra.
The spacetime and the brane worldvolume are representations of G-symmetry. 
The action is written as bilinear of
the field strength $F$.
There is a gauge symmetry generated by the Gauss law constraint with  the gauge parameter $\kappa$.
The spacetime coordinate $X_\sigma$ plays the gauge field 
while the auxiliary coordinate $X_\tau$ corresponds to $A_0$ in the usual gauge theory.
The worldvolume coordinate is denoted by $\partial=\frac{\partial}{\partial \sigma}$.
We focus on the $D$=4 E$_5$=SO(5,5) G-symmetry  case in this paper.
Representations of the G-symmetries of theories for the $D$=4 case 
are summarized  in table \ref{GsymFMTS}.
\begin{table}[h]\begin{center}
		\begin{tabular}{c|cccccc}
			theory&G-symmetry
			&\begin{tabular}{c}spacetime\\\\
				$X_\sigma$\end{tabular}	
			&\begin{tabular}{c}spacetime\\auxiliary\\
				$X_\tau$\end{tabular}
			&\begin{tabular}{c}world\\-volume\\
				$\partial$\end{tabular}
			&\begin{tabular}{c}field\\strength\\
				$F$\end{tabular}	
			&	\begin{tabular}{c}gauge\\parameter\\
				$\kappa$\end{tabular}
			\\ \hline
			F-
			&SO(5,5)&16&16'&$1 \oplus  10$ &$16 \oplus 16'$&$16 \oplus 16'$\\
			M-
			&GL(5)	
			&$5\oplus 10'\oplus 1$		&1$\oplus 10 \oplus 5' $&$1\oplus  5$
			& $1\oplus 10\oplus 5'$
			&	$1\oplus  10 \oplus  5'$\\ 
			T-
			&O(4,4)&8&0&$1 \oplus  1 $ &$8 \oplus 8$&0\\
			S-
			&GL(4)&4&0&$1 \oplus  1$& $4 \oplus 4 $&0
		\end{tabular} \label{repGofFMTS}
	\end{center}
	\caption{G-symmetries and representations of theories ($D$=4 case)}
	\label{GsymFMTS}
\end{table}
The $D$=4 F-theory is described by the 10-brane in the 16-dimensional spacetime.
The 10-dimensional chiral spinor has a bosonic $\kappa$-symmetry-like structure
similar to the Green-Schwarz superstring.
The $D$=4 F-theory in the lightcone-like gauge fixing reduces to 
the T-theory in the 4-dimensional spacetime.

The G-symmetry 
is enlarged to F-symmetry in the Lagrangian formulation where
the worldvolume Lorentz covariance is manifest.
In the usual gauge theory the G-symmetric field strengths correspond to 
the nonrelativistic electric field and  magnetic field while
the F-symmetric field strength corresponds to 
the Lorentz covariant field strength.
The constraint ${\cal V}=0$ is a worldvolume section condition.
The F-symmetries and  representations of theories 
are summarized as in the table \ref{FsymFMTS}.
\begin{table}[h]
	\begin{center}
		\begin{tabular}{c|cccccc}
			theory&F-symmetry
			&\begin{tabular}{c}space\\-time\\$X$\end{tabular}
			&\begin{tabular}{c}world\\-volume\\$\partial$\end{tabular}
			&\begin{tabular}{c}field\\strength\\
				$F$\end{tabular}
			&	\begin{tabular}{c}gauge\\parameter\\$\kappa$\end{tabular}
			&\begin{tabular}{c}cons\\-traint\\	${\cal V}=0$\end{tabular}
			\\ \hline
			F-
			&SO(6,6)&32'&12&32&32&1			\\
			M-
			&GL(6)	&$6 \oplus 20 \oplus   6'$
			&$6 \oplus  6'$
			&$1 \oplus 15 \oplus 15' \oplus  1$
			&$1 \oplus 15 \oplus  15'\oplus  1$
			&1		\\ 
			T-
			&O(4,4)SL(2)&(8,1)&(1,2)&(8,2)&0&0			\\
			S-
			&GL(4)SL(2)&(4,1)&(1,2)&(4,2)&0&0
		\end{tabular} \label{repFofFMTS}
	\end{center}
	\caption{F-symmetries and representations of theories ($D$=4 case)}
	\label{FsymFMTS}
\end{table}

The F-symmetry includes the worldvolume symmetry.
For example  the worldsheet zweibein is in the SL(2)/SO(1,1) coset parameter.
Symmetry groups for $D$=4 case are summarized as in the table \ref{symmetryFMTS}.
\begin{table}[h]\begin{center}
		\begin{tabular}{c|cccc}
			theory&F-symmetry&G-symmetry&L-symmetry&H-symmetry\\ \hline
			F-theory&SO(6,6)&SO(5,5)&SO(6;$\mathbb{C}$)&SO(5;$\mathbb{C}$)\\
			M-theory&GL(6)&GL(5)&SO(4,2)&SO(4,1)\\
			T-theory&O(4,4)SL(2)&O(4,4)&SO(3;1)$^2$SO(1,1)&SO(3,1)$^2$\\
			S-theory&GL(4)SL(2)&GL(4)&SO(3,1)SO(1,1)&SO(3,1)
		\end{tabular}
		\caption{Symmetries of theories ($D$=4 case)}
		\label{symmetryFMTS}
	\end{center}
\end{table}

The spacetime and the worldvolume vielbeins are  
elements of the coset F/L.
The number of dimensions of the coset F/L is larger than
the one of G/H by the number of   Lagrange multipliers of 
Virasoro constraints for the $p$-brane as  
\bea
\rm{dim}\displaystyle\left(\frac{\rm{F}}{\rm{L}}\right)=
\rm{dim}\displaystyle\left(\frac{\rm{G}}{\rm{H}}\right)+p+1~~~.
\eea
Coset groups and the G/H background gauge fields for the $D$=4 theories are summarized as the table \ref{repFMTS}.
\begin{table}[h]\begin{center}
		\begin{tabular}{c|ccc}
			theory&dim$\displaystyle\left(\frac{\rm{F}}{\rm{L}}\right)$
			&dim$\displaystyle\left(\frac{\rm{G}}{\rm{H}}\right)$&
			$\displaystyle\frac{\rm{G}}{\rm{H}}$~gauge fields
			~$_{(m=0,1,\cdots,3)}$
			\\ \hline
			F-theory&36&25&$g_{mn},B_{mn},C_{[\rm{RR}]}$\\
			M-theory&21&15&$g_{\hat{m}\hat{n}}~_{(\hat{m}=0,1,\cdots,4)}$\\
			T-theory&18&16&$g_{mn},B_{mn}$\\
			S-theory&12&10&$g_{mn}$
		\end{tabular} \label{gaugefieldFMTS}
	\end{center}
	\caption{Gauge fields  ($D$=4 case)}
	\label{repFMTS}
\end{table}

F-theory reduces to T, M, S-theories by reducing the spacetime dimensions or the worldvolume dimensions by the dimensional reduction or the section condition.
In this paper we reduce from the F-theory 10-brane to the M-theory 5-brane  
with preserving the SO(5,5) G-symmetry,
then the obtained M-theory 5-brane couple to SO(5,5) background gauge fields.
\par
\vskip 6mm
\section{SO(5,5) current algebra}

The $D$=4 F-theory manifests the E$_{5}$=SO(5,5) G-symmetry.
This theory reduces to the type II superstring theories in the 4 dimensional spacetime.
The type II superstring theories in 10 dimensions have 32 supersymmetries.
The supercharges have the H-symmetry index, SO(5;$\mathbb{C}$)=Sp(4;$\mathbb{C}$) index ${\cal A},\dot{{\cal A}}=1,\cdots,4$, and 
the internal space index,  SU(4) index ${\cal A}'=1,\cdots,4$.
Supercharges are ${\cal Q}_{{\cal A}{\cal A}'}$,
  $\bar{\cal Q}_{\dot{{\cal A}}}{}^{{\cal A}'}$ while translation charges are
   ${\cal P}_{{\cal A}\dot{{\cal B}}}$, 
 $\mho_{[{\cal A}{\cal B}][{\cal A}'{\cal B}']}$ and
  $\bar{\mho}_{[\dot{\cal A}\dot{\cal B}]}{}^{[{\cal A}'{\cal B}']}$.   
 The superalgebra in the $D$=4 F-theory 
  is given as follows.
\bea
\{{\cal Q}_{{\cal A}{\cal A}'},\bar{\cal Q}_{\dot{\cal B}}{}^{{\cal B}'}
\}&=&\delta_{{\cal A}'}^{{\cal B}'}{\cal P}_{{\cal A}\dot{{\cal B}}}\nn\\
\{{\cal Q}_{{\cal A}{\cal A}'},{\cal Q}_{{\cal B}}{}_{{\cal B}'}
\}&=&\mho_{[{\cal A}{\cal B}][{\cal A}'{\cal B}']}
\label{SUSY}\\
\{\bar{\cal Q}_{\dot{\cal A}}{}^{{\cal A}'},\bar{\cal Q}_{\dot{{\cal B}}}{}^{{\cal B}'}
\}&=&\bar{\mho}_{[\dot{\cal A}\dot{\cal B}]}{}^{[{\cal A}'{\cal B}']}
\nn
\eea
 The 16 translation operators ${\cal P}_{{\cal A}\dot{{\cal B}}}$ 
are decomposed into 4 momenta, 4 winding modes (NSNS charges) and 8 RR charges
($1\oplus 6\oplus 1$ for type IIA and $4\oplus 4$ for type IIB) 
in the type II superstring theories in 4 dimensions.
The rest of bosonic operators $\mho_{[{\cal A}{\cal B}][{\cal A}'{\cal B}']}$,
$\bar{\mho}_{[\dot{\cal A}\dot{\cal B}]}{}^{[{\cal A}'{\cal B}']}$
 are  
internal operators which 
 are linear combinations of
the 6-dimensional internal space  momenta $\Upsilon_{[{\cal A}'{\cal B}']}$,  $\bar{\Upsilon}^{[{\cal A}'{\cal B}']}$ and winding modes
$\digamma_{[{\cal A}{\cal B}][{\cal A}'{\cal B}']}$,  $
\digamma_{[\dot{\cal A}\dot{\cal B}]}{}^{[{\cal A}'{\cal B}']}
$ as follows.
\bea
\mho_{[{\cal A}{\cal B}][{\cal A}'{\cal B}']}
&=&C_{{\cal A}{{\cal B}}}{\Upsilon}_{[{\cal A}'{\cal B}']}+
\digamma_{[{\cal A}{\cal B}][{\cal A}'{\cal B}']}
\nn\\
\bar{\mho}_{[\dot{\cal A}\dot{\cal B}]}{}^{[{\cal A}'{\cal B}']}
&=&
C_{\dot{{\cal A}}\dot{{\cal B}}}\bar{\Upsilon}^{[{\cal A}'{\cal B}']}
+\digamma_{[\dot{\cal A}\dot{\cal B}]}{}^{[{\cal A}'{\cal B}']}
\eea

At first, we present current algebras in the SO(5,5) spinor representation.
Next, we present it in the GL(5) tensor representation, 
where reduction to M-theory is straightforward
and coupling to the 5-dimensional subspace of the 11-dimensional supergravity background is manifest.
\par\vskip 6mm

\subsection{SO(5,5) spinor representation}

The SO(5,5) current in a flat space is
 the 16-component spinor $\dd_\mu$ with $\mu=1,\cdots,16$,
 while the worldvolume is the 10-dimensional vector
with the worldvolume spacial derivative $\partial^{\underline{m}}$ for $\underline{m}=1,\cdots,10$.
The SO(5,5) current algebra in a flat space is given by 
\bea
\left[\dd_{\mu}(\sigma),\dd_{\nu}(\sigma)\right]&=&
2i\gamma_{\underline{m}\mu\nu}\partial^{\underline{m}}
\delta(\sigma-\sigma')
\label{CASO55flat}
\eea
with $\partial^{\underline{m}}\delta(\sigma)=\frac{\partial}{\partial \sigma_{\underline{m}}}\delta^{(10)}(\sigma)$.
The 10-dimensional $\gamma$ matrix is defined by
\bea
&\gamma_{(\underline{m}|\mu\rho}\gamma^{\underline{l}\rho\nu}\eta_{\underline{l}|\underline{n})}=2\eta_{\underline{m}\underline{n}}\delta_\mu^\nu~~&~\label{10dgamma}
\eea
with the SO(5,5) invariant metric $\eta_{\underline{m}\underline{n}}$.
The gamma matrix $\gamma_{\underline{m}\mu\nu}$ 
is transformed under the SO(5,5) transformation as 
\bea
&{\rm{SO}}(5,5)\ni M_{\underline{m}}{}^{\underline{n}},~\hat{M}_{\mu}{}^{\nu}~~,~~
(\hat{M}_\mu{}^{\sigma}\gamma_{\underline{n}\sigma\rho}
\hat{M}_\nu{}^{\rho})M_{\underline{m}}{}^{\underline{n}}
=\gamma_{\underline{m}\mu\nu}
~~,~~M_{\underline{l}}{}^{\underline{m}}
\eta^{\underline{lk}}M_{\underline{k}}{}^{\underline{n}}
=\eta^{\underline{mn}}~~~.&
\label{SO55gamma}
\eea 
The current algebra \bref{CASO55flat} is SO(5,5) covariant with transformations of
the spinor current and the vector derivative as
\bea
&\dd_\mu\to\hat{M}_\mu{}^\nu\dd_\nu~~,~~
\partial_{\underline{m}}\to M_{\underline{m}}{}^{\underline{n}}\partial_{\underline{n}}&\label{SO55wv}~~~.
\eea

 The 16-dimensional SO(5,5) spinor coordinate $X^\mu$ and its canonical conjugate $P_\mu$ are given by
 \bea
 \left[P_\mu(\sigma),X^\nu(\sigma')\right]&=&\frac{1}{i}\delta_\mu^\nu\delta(\sigma-\sigma')
 ~~\label{canonical}.
 \eea
The selfdual current $\dd_\mu$,
the anti-selfdual current $\tilde{\dd}_\mu$ and their algebras are 
given as
\bea
&&	{\renewcommand{\arraystretch}{1.6}
	\left\{\begin{array}{ccl}
\dd_\mu&=&P_\mu+\gamma_{\underline{m}\mu\nu}\partial^{\underline{m}}X^{\nu}
=P_\mu+\slashed{\partial}_{\mu\nu}X^\nu\nn\\
\tilde{\dd}_\mu&=&P_\mu-\gamma_{\underline{m}\mu\nu}\partial^{\underline{m}}X^{\nu}
=P_\mu-\slashed{\partial}_{\mu\nu}X^\nu
\end{array}\right.}\\\nn\\
&&	{\renewcommand{\arraystretch}{1.6}
	\left\{\begin{array}{ccl}
\left[\dd_{\mu}(\sigma),\dd_{\nu}(\sigma)\right]&=&2i\gamma_{\underline{m}\mu\nu}\partial^{\underline{m}}\delta(\sigma-\sigma')\\
\left[\dd_{\mu}(\sigma),
\tilde{\dd}_{\nu}(\sigma)\right]&=&0\label{currentflat}\\%
\left[\tilde{\dd}_{\mu}(\sigma),\tilde{\dd}_{\nu}(\sigma)\right]&=&-2i\gamma_{\underline{m}\mu\nu}\partial^{\underline{m}}\delta(\sigma-\sigma')\end{array}\right.}~~~.
\eea
Under the global SO(5,5) transformation the canonical coordinates are transformed with use of \bref{SO55gamma}
in such a way that currents are transformed as a SO(5,5) spinor 
\bea
&&	{\renewcommand{\arraystretch}{1.6}\left\{\begin{array}{ccl}
P_\mu&\to&\hat{M}_\mu{}^\nu P_\nu\nn\\
\partial^{\underline{m}}X^\mu&\to&
M^{-1}{}_{\underline{n}}{}^{\underline{m}}
(\partial^{\underline{n}}X^\nu)\hat{M}^{-1}{}_\nu{}^\mu\end{array}\right.}\nn\\
&&~~~~~~\Rightarrow
(P_\mu\pm \gamma_{\underline{m}\mu\rho}\partial^{\underline{m}}X^\rho)
\to
\hat{M}_\mu{}^\nu\left(P_\nu\pm \gamma_{\underline{m}\mu\rho}\partial^{\underline{m}}X^\rho\right)~~~.
\eea

The 10-dimensional worldvolume $\sigma_{\underline{m}}$-diffeomorphism is generated by the 
Virasoro constraint ${\cal S}^{\underline{m}}=0$, and
the $\tau$-diffeomorphism is generated by ${\cal T}$.
Closure of the Virasoro algebra requires secondary constraints, ${\cal U}^\mu=0$ 
and ${\cal V}=0$
\bea
{\cal S}^{\underline{m}}&=&\frac{1}{4}\dd_\mu \gamma^{\underline{m}\mu\nu}\dd_\nu=0\label{Virasoro}\label{Virasorospinor}\\
{\cal T}&=&\frac{1}{4}\dd_\mu \hat{\eta}^{\mu\nu}\dd_\nu=0\nn\\
{\cal U}^\mu&=&\gamma^{\underline{m}}{}^{\mu\nu}\dd_\nu
\eta_{\underline{mn}}
\partial^{\underline{n}}
=(\dd\slashed{\partial} )^\mu=0\nn\\
{\cal V}&=&\eta_{\underline{mn}}\partial^{\underline{m}}\partial^{\underline{n}}=0~~~.\nn
\eea
The constraints 
 ${\cal U}^\mu=0$ 
and ${\cal V}=0$ are the Gauss law constraints which generate infinite series of gauge symmetries similar to the $\kappa$-symmetry.
For the 16 component spacetime current $\dd_\mu$ and the 10 component 
worldvolume translation $\partial^{\underline{m}}$ which satisfy  \bref{CASO55flat},
constraints ${\cal U}^\mu=\dd\slashed{\partial}^\mu$ and ${\cal V}=\partial^{\underline{m}}\partial_{\underline{m}}$
become first class constraints. 
${\cal V}=0$ constraint is the SO(5,5) invariant constraint
sectioning the worldvolume.
${\cal S}^{\underline{m}}=0$ and ${\cal U}^\mu=0$ are SO(5,5) covariant, while  
the metric $\hat{\eta}^{\mu\nu}$ in ${\cal T}$ breaks  SO(5,5)  to SO(5;$\mathbb{C}$).

They satisfy the following algebra
\bea
\left[{\cal S}^{\underline{m}}(\sigma),{\dd}_\mu(\sigma')\right]&=&i
[2\dd_\mu\partial^{\underline{m}}-
\gamma^{\underline{m}}{}_{\mu\nu}
{\cal U}^\nu
](\sigma)\delta(\sigma-\sigma')\label{Viradd}\\
\left[{\cal T}(\sigma),{\dd}_\mu(\sigma')\right]&=&i
\hat{\eta}_{\mu\nu}{\cal U}^\nu(\sigma)\delta(\sigma-\sigma')\nn
\eea
where $\partial^{\underline{m}}$ and ${\cal U}^\mu$ act on $\delta(\sigma-\sigma')$.
The SO(5,5) indices are raised and lowered by $\eta_{\underline{mn}}$ 
and $\eta^{\underline{mn}}$.
It is also noted that $\hat{\eta}^{\mu\rho}\gamma_{\underline{m}\rho\lambda}\hat{\eta}^{\lambda\nu}=\eta_{\underline{mn}}\gamma^{\underline{n}\mu\nu}$.
The set of  Virasoro algebras is given by
\bea
\left[{\cal S}^{\underline{m}}(\sigma),{\cal S}^{\underline{n}}(\sigma')\right]&=&i
\left[2{\cal S}^{(\underline{m}}\partial^{\underline{n})}
-\frac{1}{2}\left\{\eta^{\underline{mn}}(\dd {\cal U})
-(\dd\gamma^{\underline{mn}}{\cal U})\right\}\right](\sigma)
\delta(\sigma-\sigma')\nn\\
&& +i\delta(\sigma-\sigma')
\left[\partial^{(\underline{m}}{\cal S}^{\underline{n})}
-\frac{1}{2}
({\cal U}\gamma^{\underline{mn}}\dd) 
\right]\nn\\
&=&\left[2\left({\cal S}^{\underline{n}}(\sigma)\partial^{\underline{m}}(\sigma)
+{\cal S}^{\underline{m}}(\sigma')\partial^{\underline{n}}(\sigma)\right)\right.\nn\\
&&\left.-\frac{1}{2}\left\{\eta^{\underline{mn}}(\dd {\cal U})
-(\dd\gamma^{\underline{mn}}{\cal U})\right\}(\sigma)\right]
\delta(\sigma-\sigma')\nn\\
&& +i\delta(\sigma-\sigma')
\left[\partial^{[\underline{m}}{\cal S}^{\underline{n}]}
-\frac{1}{2}
({\cal U}\gamma^{\underline{mn}}\dd) 
\right]\nn\\
\left[{\cal S}^{\underline{m}}(\sigma),{\cal T}(\sigma')\right]&=&i
  \left[4{\cal T}\partial^{\underline{m}}
  -\frac{1}{2}\dd_\mu\hat{\eta}^{\mu\nu}(\gamma^{\underline{m}}{\cal U})_\nu
\right](\sigma)\delta(\sigma-\sigma')\nn\\&&
+i\delta(\sigma-\sigma')\left[
  2\partial^{\underline{m}}{\cal T}
  -\frac{1}{2} ( {\cal U}\gamma^{\underline{m}})_{\mu}
  \hat{\eta}^{\mu\nu}\dd_\nu  \right]\nn\\
\left[{\cal T}(\sigma),{\cal T}(\sigma')\right]&=&
2i{\cal S}^{\underline{m}}\partial_{\underline{m}}(\sigma)\delta(\sigma-\sigma')
+i\delta(\sigma-\sigma')
(\partial_{\underline{m}}{\cal S}^{\underline{m}})
\label{Virasoroset}~~~\\%
\left[ {\cal S}^{\underline{m}} (\sigma),{\cal U}^\mu (\sigma')\right]
&=&i(\dd \gamma^{\underline{m}})^\mu{\cal V}(\sigma)\delta(\sigma-\sigma')
\nn\\
\left[ {\cal T} (\sigma),{\cal U}^\mu (\sigma')\right]
&=&i(\dd \hat{\eta})^\mu{\cal V}(\sigma)\delta(\sigma-\sigma')
\nn\\
\left[ {\cal U}^\mu (\sigma),{\cal U}^\nu (\sigma')\right]
&=&-
2i\slashed{\partial}^{\mu\nu}{\cal V}(\sigma)\delta(\sigma-\sigma')
\nn~~~.
\eea

The spacetime coordinate derivative of a function $\Phi(X)$ is given by
\bea
\frac{\partial}{\partial X^\mu}\Phi(X)&=&\partial_\mu\Phi(X)=i\int d^{10}\sigma'\left[ \dd_\mu(\sigma'),\Phi(X(\sigma))\right]~~~.
\eea
The worldvolume coordinate derivative of a  function 
 $\Phi\left(X (\sigma)\right)$ is given by
\bea
\frac{\partial}{\partial \sigma_{\underline{m}} } \Phi(X(\sigma))~=~
\partial^{\underline{m}} \Phi(X(\sigma))&=&i\int d^{10}\sigma' \left[{\cal S}^{\underline{m}}(\sigma'),\Phi(X(\sigma))\right]\nn\\
&=& \frac{1}{2}\dd_\mu  \gamma^{\underline{m}}{}^{\mu\nu} ~ \partial_\nu\Phi(X)~~~.
\eea

The SO(5,5) current algebra in curved backgrounds with torsion $T_{\alpha\beta}{}^\gamma$ is given as follows. 
\bea
&\left[\dd_{\alpha}(\sigma),\dd_{\beta}(\sigma')\right]~=~2i
T_{\alpha\beta}{}^{\gamma}\dd_{\gamma}\delta(\sigma-\sigma')
+i\gamma_{\underline{a}\alpha\beta}
\left(D^{\underline{a}}(\sigma)-D^{\underline{a}}(\sigma')\right)\delta(\sigma-\sigma')
\label{CASO55bg}&\\
&T_{\alpha\beta}{}^\gamma=E_{[\alpha|}{}^\mu(\partial_\mu E_{|\beta]}{}^\nu)E_{\nu}{}^{\gamma}
-\frac{1}{2}(E_{[\alpha|}{}^\mu
\gamma_{\underline{a}\mu\nu}
\partial_\lambda E_{|\beta]}{}^\nu)
\gamma^{\underline{a}\lambda\rho}
E_\rho{}^\gamma&\nn
\eea
The curved space current $\dd_\alpha$, the flat space current $\dd_\mu$ and 
the curved space worldvolume derivative $D^{\underline{a}}$, the flat worldvolume derivative $\partial^{\underline{m}}$  are related by
 the spacetime vielbein $E_\alpha{}^\mu$ and the worldvolume vielbein  ${\cal E}_{\underline{m}}{}^{\underline{a}}$ 
as
\bea
&\dd_\alpha=E_\alpha{}^\mu\dd_\mu~~,~~
D^{\underline{a}}={\cal E}_{\underline{m}}{}^{\underline{a}}\partial^{\underline{m}}~~~.&
\eea
The spacetime vielbein $E_\alpha{}^\mu\in$ SO(5,5)/SO(5;C) 
and the worldvolume vielbein
relate the curved background indices $\mu$, $\underline{m}$ 
and the flat space indices $\alpha$, $\underline{a}$ as
\bea
E_\alpha{}^\mu\gamma_{\underline{m}\mu\nu}E_\beta{}^\nu
{\cal E}_{\underline{a}}{}^{\underline{m}}=
\gamma_{\underline{a}\alpha\beta}~~,~~
{\cal E}_{\underline{a}}{}^{\underline{m}}\eta_{\underline{m}\underline{n}}{\cal E}_{\underline{b}}{}^{\underline{n}}=\eta_{\underline{a}\underline{b}}\label{conditionSO55}
\eea
The gauge transformation of $E_\alpha{}^\mu$ is given by
\bea
\delta_\lambda E_\alpha{}^\mu\dd_\mu(\sigma)&=&
i\int d\sigma' \left[\lambda^\nu\dd_\nu(\sigma'),E_\alpha{}^\mu\dd_\mu(\sigma)\right]\nn\\
\delta_\lambda E_\alpha{}^\mu&=&\lambda^\nu\partial_\nu
E_\alpha{}^\mu
-E_\alpha{}^\nu\partial_\nu \lambda^\mu
+(E_\alpha\gamma^{\underline{a}}\partial_\nu \lambda) \gamma_{\underline{a}}{}^{\nu\mu}~~~.
\eea

In curved backgrounds 
 the $\tau$ component of Virasoro constraint ${\cal T}=0$  \bref{Virasoro}  is generalized as 
\bea
{\cal T}&=&\frac{1}{4}\dd_\mu{G}^{\mu\nu}\dd_\nu=
\frac{1}{4}\dd_\alpha\hat{\eta}^{\alpha\beta}\dd_\beta\\
{G}^{\mu\nu}&=&E_\alpha{}^\mu \hat{\eta}^{\alpha\beta}E_\beta{}^\nu\nn~~~.
\eea
 The spacetime gauge field $G^{\mu\nu}$ is parametrized by elements of the coset SO(5,5)/SO(5;$\mathbb{C}$).
Spacial components of Virasoro constraints ${\cal S}^{\underline{m}}$ in \bref{Virasoro} is inert in curved backgrounds as
\bea
\frac{1}{4}\dd_\alpha\gamma^{\underline{a}\alpha\beta}\dd_\beta 
=\frac{1}{4}{\cal E}_{\underline{m}}{}^{\underline{a}}(\dd_\mu\gamma^{\underline{m}\mu\nu}
\dd_\nu)={\cal E}_{\underline{m}}{}^{\underline{a}}{\cal S}^{\underline{m}}~~~.
\eea
Background independence of the spacial components of Virasoro constraints 
makes possible to impose  as the section conditions on fields.
This is the same property with the T-theory. 

\par
\vskip 6mm

\subsection{GL(5) tensor representation}

The SO(5,5) current algebra in the GL(5) tensor representation was obtained 
in the M5 brane Hamiltonian\cite{Hatsuda:2013dya} from 
the PST action\cite{Pasti:1997gx}. The M5-brane is a 11-dimensional supergravity solution which is described by the spacetime coordinate $x^m(\sigma)$, $m=0,1,\cdots,10$, the second rank 
selfdual gauge field $A_{ij}(\sigma)$, $i=1,\cdots,5$  and their canonical conjugates $p_m(\sigma)$, $E^{ij}(\sigma)$.
The currents are the vector, the 2-rank tensor corresponding to the M2 brane charge and
the 5-rank tensor corresponding to the M5 brane charge.
The $\tau$-diffeomorphism constraint ${\cal T}=\frac{1}{2}p_m{}^2+\cdots=0$ is written in bilinear of currents.
On the other hand the 5-dimensional worldvolume diffeomorphism constraints are 
given as ${\cal H}_i=p_n\partial_i x^m+\frac{1}{2}E^{j_1j_2}\partial_{[i}A_{j_1j_2]}=0$.
Multiplying the pullback matrices $\partial_i x^m$ and $E^{ij}\partial_jx^m$ on ${\cal H}_i=0$ gives 
bilinears of currents 
${\cal S}^m={\cal S}^{m_1\cdots m_4}=0$ as Virasoro constraints.

In the 5-dimensional subspace where the 11-dimensional space is compactified on a 5-dimensional torus
the currents  combine into the 16 dimensional SO(5,5) spinor representation.
The reducible set of Virasoro constraints  ${\cal S}^m={\cal S}^{m_1\cdots m_4}=0$ 
become ${\cal S}^m=\bar{\cal S}_{m}=0$ with $m=1,\cdots, 5$
which is the $5+\bar{5}$ vector representation of the SO(5,5).
All these constraints ${\cal T}$, ${\cal S}^m$ and $\bar{\cal S}_m$ satisfy
a closed algebra with secondary constraints.

The SO(5,5) current algebra is written in terms of the 16-dimensional spacetime currents $\dd_M=(\dd_m,~\dd^{m_1m_2},~\bar{\dd})$ 
as the GL(5) decomposition of 16 component SO(5,5) spinor current $16 \to 5 \oplus 10 \oplus 1$.
The commutator of the spinor currents gives the vector,
so the worldvolume is 10-dimensional space with $\partial^{\underline{m}}=(\partial^m,~\bar{\partial}_{m})$ with $
\underline{m}=1,\cdots, 10$, and $m=1,\cdots,5$.
The SO(5,5) current algebra in the GL(5) tensor representation is given 
with the 10-dimensional gamma matrices $\rho_{\underline{l}MN}$
and  $\rho^{\underline{l}MN}$ based on \cite{Hatsuda:2013dya}
by 
\bea
&\left[\dd_M(\sigma),\dd_N(\sigma')\right]~=~2i\rho_{{\underline{m}
	}MN}\partial^{\underline{m}}\delta(\sigma-\sigma')\label{SO55GL5tensor}&~~~\\
&\rho_{(\underline{l}|ML}~\rho^{\underline{n}LN}~
\eta_{\underline{n}|\underline{k})}
~=~
\eta_{\underline{lk}}\delta_M^N&\label{rho}~~~.\nn
\eea
The O(5,5) invariant metric $\eta_{\underline{mn}}$
is given by
\bea
&&~~~~~~~~~_n~~~~~^n\nn\\
  \eta_{\underline{mn}}&=&
  \begin{array}{c} _m \\^m 
  \end{array}
  \left(\begin{array}{cc}
  	0&\delta_m^n\\
  	\delta_n^m&0
  \end{array}\right)~~~.
\eea
Concrete expression of $\rho_{\underline{l}MN}$ with arbitrary parameters 
$s^{\underline{l}}=(s^l,~\bar{s}_l)$ 
and $s_{\underline{l}}=\eta_{\underline{lm}}s^{\underline{m}}=(\bar{s}_l,~{s}^l)$ 
is given by
\bea
&&~~~~~~~~~~~~~~~~_n~~~~~~~~~~~~^{n_1n_2}~~~~~~~~\cdot~~~~~~~~~~~
\nn\\~~
\rho_{\underline{l}MN}s^{\underline{l}}&=&
\begin{array}{c} _m \\^{m_1m_2}  \\	\cdot
\end{array}
\left(\begin{array}{ccc}
0&\delta_m^{[n_1}s^{n_2]}&\bar{s}_m\\
\delta_n^{[m_1}s^{m_2]}&\epsilon^{m_1m_2n_1n_2k}\bar{s}_{k}&0\\
\bar{s}_n&0&0
\end{array}\right)\nn\\
\label{rhorho}\\
&&~~~~~~~~~~~~~~~~^n~~~~~~~~~~~~_{n_1n_2}~~~~~~~~\cdot~~~~~~~~~~~
\nn\\~~
\rho^{\underline{l}MN}s_{\underline{l}}&=&
\begin{array}{c} ^m \\_{m_1m_2}  \\	\cdot
\end{array}
\left(\begin{array}{ccc}
	0&\delta^m_{[n_1}\bar{s}_{n_2]}&{s}^m\\
	\delta^n_{[m_1}\bar{s}_{m_2]}&\epsilon_{m_1m_2n_1n_2k}{s}^{k}&0\\
	{s}^n&0&0
\end{array}\right)\nn
\eea
 $\rho^{\underline{l}MN}$ satisfies the same  SO(5,5) transformation \bref{SO55gamma}.
Under the SO(5,5) transformation  it is transformed as
\bea
&&{\rm{SO}}(5,5)\ni 
M_{\underline{m}}{}^{\underline{n}}=\delta_{\underline{m}}{}^{\underline{n}}+\delta M_{\underline{m}}{}^{\underline{n}},
~\hat{M}_{M}{}^{N}=\delta_{M}{}^{N}+
\delta \hat{M}_{M}{}^{N}\nn\\
&&(\rho^{\underline{l}ML}s_{\underline{l}})\delta \hat{M}_{L}{}^{N}+
\delta \hat{M}_{L}{}^{M}(\rho^{\underline{l}LN}s_{\underline{l}})+
\rho^{\underline{l}MN}\delta {M}_{\underline{l}}{}^{\underline{n}}s_{\underline{n}}=0
\label{SO55M}\\
&&\eta^{\underline{ml}}\delta {M}_{\underline{l}}{}^{\underline{n}}
+\delta {M}_{\underline{l}}{}^{\underline{m}}\eta^{\underline{ln}}=0\nn
\eea
where infinitesimal SO(5,5) matircies are given by
\bea
&&~~~~~~~~~~~~~~~~~~^n~~~~~~~~~~~~~~~~~~_{n_1n_2}~~~~~~~~~~~~~~~~~~\cdot~~~~~~~~~~~
\nn\\~~
\delta \hat{M}_{M}{}^{N}&=&
\begin{array}{c} _m \\^{m_1m_2}  \\	\cdot
\end{array}
\left(\begin{array}{ccc}
\alpha_m{}^n-\frac{\hat{\alpha}}{2}\delta_m^n&
-\tilde{\gamma}_{mn_1n_2}&0\\
-\tilde{\beta}^{m_1m_2n}&\frac{\hat{\alpha}}{2}\delta_{n_1}^{[m_1}
\delta_{n_2}^{m_2]}-
\delta_{[n_1}^{[m_1}\alpha_{n_2]}{}^{m_2]}&\gamma^{m_1m_2}\\
0&\beta_{n_1n_2}&\frac{\hat{\alpha}}{2}
\end{array}\right)\nn\\
\label{Mst}\\
&&~~~~~~~~~~~^n~~~~~~~~~_n
~~~~~~~~~~~~~~~~~~~~~~~
\nn\\~~
\delta {M}_{\underline{m}}{}^{\underline{n}}&=&
\begin{array}{c} _m \\^m
\end{array}
\left(\begin{array}{cc}
	\alpha_m{}^n&\beta_{mn}\\
	-\gamma^{mn}&-\alpha_n{}^m
\end{array}\right)\nn~~~
\nn~~~\\
\rm{with}&&
\tilde{\beta}^{m_1m_2m_3}=\frac{1}{2}\epsilon^{m_1\cdots m_5}\beta_{m_4m_5}~~,~~
\tilde{\gamma}_{m_1m_2m_3}=\frac{1}{2}\epsilon_{m_1\cdots m_5}\gamma^{m_4m_5}~~~.\nn
\eea

The GL(5) tensor coordinate $X^M=(X^m,~X_{m_1m_2},~\bar{X})$ and its canonical conjugate $P_M=(P_m,~P^{m_am_2},~\bar{P})$
are introduced by
\bea
\left[P_M(\sigma),X^N(\sigma')\right]&=&\frac{1}{i}\delta^N_M\delta(\sigma-\sigma')\label{XPGL5}~~~.
\eea
The selfdual and anti-selfdual currents,$\dd_M$ and $\tilde{\dd}_M$, together with 
their current algebras 
are given by
\bea
&&	{\renewcommand{\arraystretch}{1.6}\left\{\begin{array}{ccl}
	\dd_M&=&P_M+\rho_{\underline{m}MN}\partial^{\underline{m}}X^{N}
	\nn\\
	\tilde{\dd}_M&=&P_M-\rho_{\underline{m}MN}\partial^{\underline{m}}X^N
\end{array}\right.}\\\nn\\
&&	{\renewcommand{\arraystretch}{1.6}\left\{\begin{array}{ccl}
	\left[\dd_M(\sigma),\dd_N(\sigma)\right]&=&2i\rho_{\underline{m}MN}\partial^{\underline{m}}\delta(\sigma-\sigma')\\
	\left[\dd_M(\sigma),
	\tilde{\dd}_N(\sigma)\right]&=&0
	\\
	\left[\tilde{\dd}_M(\sigma),\tilde{\dd}_N(\sigma)\right]&=&-2i\rho_{\underline{m}MN}\partial^{\underline{m}}\delta(\sigma-\sigma')\end{array}\right.}~~~.\label{currentflatGL5}
\eea
The selfdual currents and their algebra in components 
are given as
\bea
&&	{\renewcommand{\arraystretch}{1.6}\left\{\begin{array}{ccl}
	\dd_m&=& P_m+\partial^n X_{mn}+\bar{\partial}_m\bar{X}\\
	\dd^{m_1m_2}&=& P^{m_1m_2}+\partial^{[m_2} X^{m_1]}+
	\frac{1}{2}\epsilon^{m_1 \cdots m_4l}\bar{\partial}_lX_{m_3 m_4}\\
	\bar{\dd}&=& \bar{P}+\bar{\partial}_m X^{m}\end{array}\right.}~~~
\nn
\\
&&	{\renewcommand{\arraystretch}{1.6}\left\{\begin{array}{rcl}
\left[\dd_m(\sigma),\dd^{n_1n_2}(\sigma')\right]&=&2i\delta_m^{[n_1}\partial^{n_2]}\delta(\sigma-\sigma')\\
\left[\dd_m(\sigma),\bar{\dd}(\sigma')\right]&=&2i\bar{\partial}_m\delta(\sigma-\sigma')\\
\left[\dd^{m_1m_2}(\sigma),\dd^{n_3n_4}(\sigma')\right]&=&2i\epsilon^{m_1m_2n_3n_4l}\bar{\partial}_l
\delta(\sigma-\sigma')\end{array}\label{CA3}
\right.}
\eea

The set of Virasoro constraints in \bref{Virasoro} is rewritten as 
\bea
{\cal S}^{\underline{m}}&=&\frac{1}{4}\dd_M\rho^{\underline{m}MN}\dd_N=0
\label{VirasoroGauss}\\
{\cal T}&=&\frac{1}{4}\dd_M\hat{\eta}^{MN}\dd_N=0\nn\\
{\cal U}^M&=&\rho^{\underline{m}MN}\eta_{\underline{mn}}\dd_N\partial^{\underline{n}}=0\nn\\
{\cal V}&=&\eta_{\underline{mn}}\partial^{\underline{m}}{\partial}^{\underline{n}}=0\nn
\eea
with the  SO(5;$\mathbb{C}$) invariant metric
\bea
&&~~~~~~~~~~~~~~~~^n~~~~~~~~~~~~_{n_1n_2}~~~~~~~~\cdot~~~~~~~~~~~
\nn\\~~
\hat{\eta}^{MN}&=&
\begin{array}{c} ^m \\_{m_1m_2}  \\	\cdot
\end{array}
\left(\begin{array}{ccc}
	\eta^{mn}&0&0\\
0&\eta_{n_1[m_1}\eta_{m_2]n_2}&0\\
	0&0&1
\end{array}\right)\label{hateta1}~~~.
\eea
The Virasoro constraints in components are given by
\bea
&&	{\renewcommand{\arraystretch}{1.6}\left\{\begin{array}{rcl}
{\cal S}^m&=&\frac{1}{2}\dd_n \dd^{nm}=0\\
\bar{\cal S}_{m}&=&\frac{1}{2}\left[\dd_m\bar{\dd}+\frac{1}{8}
\epsilon_{mm_1\cdots m_4}\dd^{m_1m_2}\dd^{m_3m_4}\right]
=0\\
{\cal T}&=&\frac{1}{4}\left[\dd_m\eta^{mn}\dd_n+\frac{1}{2} \dd^{m_1m_2}\eta_{m_1n_1}\eta_{m_2n_2}\dd^{n_1n_2}
+\bar\dd^2\right]
=0
\end{array}\right.}\label{Virasoro2}~~~.\\
&&	{\renewcommand{\arraystretch}{1.6}\left\{\begin{array}{rcl}
{\cal U}^m&=&\bar{\dd}\partial^m+\dd^{ml}\bar{\partial}_l=0\\
{\cal U}_{m_1m_2}&=&\dd_{[m_1}\bar{\partial}_{m_2]}+\frac{1}{2}
\epsilon_{m_1\cdots m_5} \dd^{m_3m_4}\partial^{m_5}
=0 \\
\bar{\cal U}&=&\dd_m \partial^m=0\label{Ucon}~~~\\
{\cal V}&=&2\partial^m\bar{\partial}_m=0
\end{array}\right.}~~~.
\eea
The Virasoro constraint ${\cal S}^{\underline{m}}$ generates
the shift of the worldvolume coordinate on the current $\dd_M$ as
\bea
\left[{\cal S}^{\underline{m}}(\sigma),\dd_{M}(\sigma')\right]
=i\left[2\dd_M\partial^{\underline{m}}
-\eta^{\underline{ml}}\rho_{\underline{l}MN}{\cal U}^N
\right]\delta(\sigma-\sigma')
\eea
The Virasoro algebra in the GL(5) tensor representation is
the same as \bref{Virasoroset} by replacing the $\gamma^{\underline{m}\mu\nu}$-matrices
with $\rho^{\underline{m}MN}$ in \bref{rhorho}
\bea
\left[{\cal S}^{\underline{m}}(\sigma),{\cal S}^{\underline{n}}(\sigma')\right]&=&i
\left[2{\cal S}^{(\underline{m}}\partial^{\underline{n})}
-\frac{1}{2}\left\{\eta^{\underline{mn}}(\dd {\cal U})
-(\dd\rho^{\underline{mn}}{\cal U})\right\}\right](\sigma)
\delta(\sigma-\sigma')\nn\\
&& +i\delta(\sigma-\sigma')
\left[\partial^{(\underline{m}}{\cal S}^{\underline{n})}
-\frac{1}{2}
({\cal U}\rho^{\underline{mn}}\dd) 
\right]\nn\\
\left[{\cal S}^{\underline{m}}(\sigma),{\cal T}(\sigma')\right]&=&i
\left\{4{\cal T}\partial^{\underline{m}}
-\frac{1}{2}\dd_M\hat{\eta}^{MN}(\rho^{\underline{m}}{\cal U})_N
\right\}(\sigma)\delta(\sigma-\sigma')\nn\\&&
+i\delta(\sigma-\sigma')\left\{
2\partial^{\underline{m}}{\cal T}
-\frac{1}{2} ( {\cal U}\rho^{\underline{m}})_M
\hat{\eta}^{MN}\dd_N  \right\}\nn\\
\left[{\cal T}(\sigma),{\cal T}(\sigma')\right]&=&
2i{\cal S}^{\underline{m}}\partial_{\underline{m}}(\sigma)\delta(\sigma-\sigma')
+i\delta(\sigma-\sigma')
(\partial_{\underline{m}}{\cal S}^{\underline{m}})\nn\\\label{Virasoroset2}
~~~\\
\left[ {\cal S}^{\underline{m}} (\sigma),{\cal U}^M (\sigma')\right]
&=&i(\dd \rho^{\underline{m}})^M{\cal V}(\sigma)\delta(\sigma-\sigma')
\nn\\
\left[ {\cal T} (\sigma),{\cal U}^M (\sigma')\right]
&=&i(\dd \hat{\eta})^M{\cal V}(\sigma)\delta(\sigma-\sigma')
\nn\\
\left[ {\cal U}^M (\sigma),{\cal U}^N (\sigma')\right]
&=&-
2i\slashed{\partial}^{MN}{\cal V}(\sigma)\delta(\sigma-\sigma')
\nn~~~
\eea
with $(\rho^{\underline{mn}})^{M}{}_N=
\frac{1}{2}\rho^{[\underline{m}|ML}\rho_{\underline{l}LN}\eta^{\underline{l}|\underline{n}]}$. 
The GL(5) tensor expression of the above relation is as follows.
\bea
	{\renewcommand{\arraystretch}{1.6}\left\{\begin{array}{rcl}
\left[{\cal S}^m(\sigma),\dd_n(\sigma')\right]&=&i
\left[
2\dd_n\partial^m-\delta_n^m\bar{\cal U}
\right](\sigma)
\delta(\sigma-\sigma')\\
\left[{\cal S}^m(\sigma),\dd^{n_1n_2}(\sigma')\right]&=&i\left[
2\dd^{n_1n_2}\partial^m-\frac{1}{2}\epsilon^{mn_1n_2n_3n_4}{\cal U}_{n_3n_4}\right](\sigma)
\delta(\sigma-\sigma')\\
\left[{\cal S}^m(\sigma),\bar{\dd}(\sigma')\right]&=&i\left[
2\bar{\dd}\partial^m-{\cal U}^m\right](\sigma)
\delta(\sigma-\sigma')\\
\left[\bar{\cal S}_m(\sigma),\dd_n(\sigma')\right]&=&i\left[
2\dd_n\bar{\partial}_m+{\cal U}_{mn}\right](\sigma)
\delta(\sigma-\sigma')\\
\left[\bar{\cal S}_m(\sigma),\dd^{n_1n_2}(\sigma')\right]&=&i\left[
2\dd^{n_1n_2}(\sigma)\bar{\partial}_m+\delta_m^{[n_1}{\cal U}^{n_2]}\right](\sigma)
\delta(\sigma-\sigma')\\
\left[\bar{\cal S}_m(\sigma),\bar{\dd}(\sigma')\right]&=&2i
\bar{\dd}(\sigma)\bar{\partial}_m\delta(\sigma-\sigma')\end{array}\right.}\label{Sdd}
\eea
The Virasoro algebra in the GL(5) tensor expression is given by
\bea
&&	{\renewcommand{\arraystretch}{1.2}
\begin{array}{rcl}
\left[{\cal S}^m(\sigma),{\cal S}^n(\sigma')\right]&=&i\left[
2{\cal S}^{(m}\partial^{n)}
-\frac{1}{2}(\dd^{mn}\bar{\cal U}-\frac{1}{2}\epsilon^{mnl_1l_2l_3}
\dd_{l_1}{\cal U}_{l_2l_3})\right](\sigma)\delta(\sigma-\sigma')\nn\\
&&
+i\delta(\sigma-\sigma')\left[
(\partial^{(m}{\cal S}^{n)})-
\frac{1}{2}(
\bar{\cal U}\dd^{mn}-\frac{1}{2}\epsilon^{mnl_1l_2l_3}{\cal U}_{l_1l_2}\dd_{l_3})\right]
\nn\\
\left[{\cal S}^m(\sigma),\bar{\cal S}_n(\sigma')\right]&=&
i\left[
2({\cal S}^m\bar{\partial}_n+\bar{\cal S}_n\partial^m)
-\frac{1}{4}\delta_n^m
(\dd_l{\cal U}^l+3\bar{\dd}\bar{\cal U}
+\frac{3}{2}\dd^{l_1l_2}{\cal U}_{l_1l_2})\right.\nn\\
&&\left.
-\frac{1}{2}(\dd_n{\cal U}^m-\dd^{ml}{\cal U}_{nl})
\right](\sigma)\delta(\sigma-\sigma')\nn\\
&&
+i\delta(\sigma-\sigma')
\left[
(\partial^m\bar{\cal S}_n+\bar{\partial}_n{\cal S}^m)
+\frac{1}{4}\delta_n^m
({\cal U}^l\dd_l-\bar{\cal U}\bar{\dd}-\frac{1}{2}{\cal U}_{\l_1l_2}\dd^{l_1l_2})\right.\nn\\
&&\left.-\frac{1}{2}({\cal U}^m\dd_n
-{\cal U}_{nl}\dd^{ml})
\right]
\nn\\
\left[\bar{\cal S}_{m}(\sigma),\bar{\cal S}_n(\sigma')\right]&=&
i\left[
2\bar{\cal S}_{(m}\bar{\partial}_{n)}
+\frac{1}{2}(\bar{\dd}{\cal U}_{mn}
-\frac{1}{2}\epsilon_{mnl_1l_2l_3}\dd^{l_1l_2}{\cal U}^{l_3})
\right](\sigma)\delta(\sigma-\sigma')\nn\\
&&
+i\delta(\sigma-\sigma')
\left[\bar{\partial}_{(m}\bar{\cal S}_{n)}
+\frac{1}{2}(
{\cal U}_{mn}\bar{\dd}
-\frac{1}{2}\epsilon_{mnl_1l_2l_3}{\cal U}^{l_1}\dd^{l_2l_3}
)
\right]\end{array}}
\nn
\\
&&	{\renewcommand{\arraystretch}{1.2}
\begin{array}{ccl}
\left[{\cal S}^m(\sigma),{\cal T}(\sigma')\right]&=&
i\left[4{\cal T}\partial^m
-\frac{1}{2}(
\eta^{mk}\dd_k\bar{\cal U}+\frac{1}{4}\epsilon^{mm_1\cdots m_4}
\eta_{m_1k_2}\eta_{m_2k_2}\dd^{k_1k_2}{\cal U}_{m_3m_4}\right.\nn\\
&&\left.+\bar{\dd}{\cal U}^m)\right](\sigma)\delta(\sigma-\sigma')\nn\\
&&+i\delta(\sigma-\sigma')\left[
2\partial~m{\cal T}-\frac{1}{2}(
\eta^{mk}\bar{\cal U}\dd_k
+{\cal U}^m\bar{\dd}\right.\nn\\
&&\left.+\frac{1}{4}\epsilon^{mm_1\cdots m_4}\eta_{m_1k_1}\eta_{m_2k_2}
{\cal U}_{m_3m_4}\dd^{k_1k_2})\right](\sigma)\delta(\sigma-\sigma')\nn\\
\left[\bar{\cal S}_m(\sigma),{\cal T}(\sigma')\right]&=&
i\left[4{\cal T}\bar{\partial}_m-\frac{1}{2}
(\eta^{kn}\dd_k{\cal U}_{nm}
+\eta_{k_1l}\eta_{k_2m}\dd^{k_1k_2}{\cal U}^l)\right](\sigma)\delta(\sigma-\sigma')\nn\\
&&+i\delta(\sigma-\sigma')
\left[2\bar{\partial}_m{\cal T}
-\frac{1}{2}(\eta^{kn}{\cal U}_{km}\dd_n
+\eta_{k_1l}\eta_{k_2m}{\cal U}l\dd^{k_1k_2})\right]\nn\\
\left[{\cal T}(\sigma),{\cal T}(\sigma')\right]&=&
2i({\cal S}^m\bar{\partial}_m
+\bar{\cal S}_m{\partial}^m)(\sigma)\delta(\sigma-\sigma')
+i\delta(\sigma-\sigma')
(\partial^m\bar{\cal S}_m+\bar{\partial}_m{\cal S}^m)
\end{array}}
\\
&&\label{VirasoroT}
\eea

In order to couple to the 11-dimensional supergravity background
the 5-dimensional indices are converted into the 11-dimensional tensor indices as
\bea
P^{m_1\cdots m_5}=\epsilon^{m_1\cdots m_5}\bar{P}~&,&~
X_{m_1\cdots m_5}=\epsilon_{m_1\cdots m_5}\bar{X}~\nn\\
\dd^{m_1\cdots m_5}=\epsilon^{m_1\cdots m_5}\bar{\dd}~&,&~
\tilde{\dd}^{m_1\cdots m_5}=\epsilon^{m_1\cdots m_5}\bar{\tilde{\dd}}~\label{pseudo}\\
\partial^{m_1\cdots m_4}=\epsilon^{m_1\cdots m_4 l}\bar{\partial}_l~&,&
~{\cal S}^{m_1\cdots m_4}=\epsilon^{m_1\cdots m_4 l}\bar{\cal S}_l\nn~~~.
\eea
Currents in the GL(5) tensor representation 
coupled to the 5-dimensional subspace of
the 11-dimensional supergravity background are given as   
\bea
&\dd_A=E_A{}^M\dd_M&~~\nn\\
&E_A{}^M\rho_{\underline{m}MN} E_B{}^N
{\cal E}_{\underline{a}}{}^{\underline{m}}=\rho_{\underline{a}AB}~~,~~
{\cal E}_{\underline{a}}{}^{\underline{m}}\eta_{\underline{mn}} {\cal E}_{\underline{a}}{}^{\underline{m}}=\eta_{\underline{ab}}\nn\\
&
E_A{}^M=\left(
\begin{array}{ccc}
	e_a{}^m&e_a{}^nC^{[3]}_{nm_1m_2}&-e_{a}{}^n
	\frac{1}{4!}
	C^{[3]}_{n[m_1m_2}C^{[3]}_{m_3m_4m_5]}
	\\
	0&e_{m_1}{}^{a_1}e_{m_2}{}^{a_2}&-\frac{1}{3!}e_{[m_1}{}^{a_1}e_{m_2}{}^{a_2}C^{[3]}_{m_3m_4m_5]}\\
	0&0&e_{[m_1}{}^{a_1}\cdots e_{m_5]}{}^{a_5}
\end{array}\right)\label{EEC3}
&	
\eea
Under the SO(5,5) transformation in  ${\cal T}=0$,  
$e_a{}^m$ and $C_{m_1m_2m_3}^{[3]}$ are trasnformed fractional linearly. The $\tau$ component of the Virasoro constraints ${\cal T}=0$ in a curved background is given by 
\bea
{\cal T}&=&\frac{1}{4}
{\dd}_M{G}^{MN}
{\dd}_N=
\frac{1}{4}\dd_A\hat{\eta}^{AB}\dd_B\\
{G}^{MN}&=&E_A{}^M \hat{\eta}^{AB}E_\beta{}^N\nn
\eea 
where $\hat{\eta}^{AB}$ is the same matrix as $\hat{\eta}^{MN}$ in \bref{hateta1}.

\par\vskip 6mm

\section{F-theory 10-brane actions}

The F-theory SO(5,5) current algebras \bref{currentflat} or
\bref {currentflatGL5}
are realized on the 10-brane worldvolume which we call F10-brane for short from now on.
The Hamiltonian is given by linear combinations of a set of Virasoro constraints \bref{Virasorospinor} or \bref{VirasoroGauss}
which are written in terms of the selfdual currents.  
In order to construct the worldvolume covariant Lagrangian we include
the ones for the anti-selfdual currents.
At first, we review how to construct the worldsheet covariant action by
using the double zweibein method \cite{{Hatsuda:2018tcx},{Hatsuda:2019xiz}}. 
Then we propose actions for the F10-brane with both the SO(5,5) symmetric Hamiltonian formulation and the SO(6,6) symmetric Lagrangian formulation.
\par
\vskip 6mm
\subsection{Double zweibein formulation in T-theory}

The physical current in the T-theory is the selfdual current
which is chiral in the doubled space.
For the doubled space coordinates $X=(x,y)$ 
the auxiliary coordinate $y$ is introduced with the selfduality condition;
 the anti-selfdual current is zero $\partial_mx-\epsilon_{mn}\partial^n y=0$.
We impose the selfduality condition as the first class constraint
by squaring of it.
The action contains both the selfdual current and the anti-selfdual current leading to the worldsheet covariant action.

The double zweibein formulation of T-theory is given by O($D,D$) current algebras for the selfdual and the anti-selfdual currents, $\dd_M$ and  $\tilde{\dd}_{M}$.
These currents are written in terms of
the O($D,D$) coordinates, $X^M$ and $P_M$ with $M=1,\cdots, 2D$,  as  
\bea
&&\left[P_M(\sigma),X^N(\sigma')\right]=\frac{1}{i}\delta_M^N\partial_\sigma\delta(\sigma-\sigma')\nn\\
&&	{\renewcommand{\arraystretch}{1.6}\left\{\begin{array}{l}
\dd_M=P_M+\partial_\sigma X^N\eta_{NM}\\
\tilde{\dd}_M=P_M-\partial_\sigma X^N\eta_{NM}
\end{array}\right.}\nn\\
&&	{\renewcommand{\arraystretch}{1.6}\left\{\begin{array}{l}
	\left[\dd_M(\sigma),\dd_N(\sigma')\right]
	=2i\eta_{MN}\partial_\sigma\delta(\sigma-\sigma')\nn\\
	\left[{\dd}_M(\sigma),\tilde{\dd}_N(\sigma')\right]
	=0\\
	\left[\tilde{\dd}_M(\sigma),\tilde{\dd}_N(\sigma')\right]
	=-2i\eta_{MN}\partial_\sigma\delta(\sigma-\sigma')
\end{array}\right.}
\eea
with $\partial_\sigma\delta(\sigma)=\frac{\partial}{\partial \sigma}\delta^{(1)}(\sigma)$.
Virasoro constraints in terms of the selfdual current
and the anti-selfdual currents are given by 
\bea
	{\renewcommand{\arraystretch}{1.6}\left\{
\begin{array}{l}
	{\cal T}=\frac{1}{4}\dd_M\hat{\eta}^{MN}\dd_N\\
	{\cal S}=\frac{1}{4}\dd_M\eta^{MN}\dd_N
	\end{array}
\right.}
~~,~~	{\renewcommand{\arraystretch}{1.6}\left\{
\begin{array}{l}
	\tilde{\cal T}=\frac{1}{4}\tilde{\dd}_M\hat{\eta}^{MN}\tilde{\dd}_N\\
	\tilde{\cal S}=\frac{1}{4}\tilde{\dd}_M\eta^{MN}\tilde{\dd}_N
\end{array}
\right.}
\eea
where $\hat{\eta}^{MN}$ and $\eta^{MN}$ are the doubled Minkowski metric and the O($D,D$) invariant metric. 
The Hamiltonian form action is given by
\bea
I&=&\int d\tau d\sigma ~L~~,~~L=\dot{X}^MP_M-H\nn\\
H&=&g{\cal T}+s{\cal S}+\tilde{g}\tilde{\cal T}+\tilde{s}\tilde{\cal S}
\label{33}\\
&=&\frac{1}{4}\dd_M(g\hat{\eta}+s\eta)^{MN}\dd_N+
\frac{1}{4}\tilde{\dd}_M(\tilde{g}\hat{\eta}+\tilde{s}\eta)^{MN}\tilde{\dd}_N\nn
\eea
with Lagrange multipliers $g$, $s$, $\tilde{g}$, $\tilde{s}$.
After the Legendre transformation the obtained Lagrangian is given by
\bea
L&=&\varphi J_+\left\{
(g+\tilde{g})\hat{\eta}-(s+\tilde{s})\eta\right\}J_-\label{LRgs}\\
&&\left\{\begin{array}{l}
	J_+=\dot{X}+(\tilde{g}\hat{\eta}+\tilde{s}\eta)\partial_\sigma X\\
	J_-=\dot{X}-({g}\hat{\eta}+{s}\eta)\partial_\sigma X
	\end{array}\right.\nn\\
&&\varphi=
\left[{(g+\tilde{g})^2-(s+\tilde{s})^2}\right]^{-1}\nn
\eea
with $\dot{X}=\partial_\tau X$.

The doubled coordinate has the left moving and the right moving components $X^{{M}}=(X^{\overline{M}},X^{\underline{M}})$
with ${\overline{M}},{\underline{M}}=1,\cdots,D$.
The $D$-dimensional right moving subscript $\underline{M}$
is only used in these two paragraphs,
and should not be confused  with the enlarged dimension subscript.
$D$-dimensional metrics are $\hat{\eta}_{MN}=$ diag $(\eta_{\overline{M}\overline{N}},
\eta_{\underline{M}\underline{N}})$ and the O($D,D$) invariant metric
 $\eta_{MN}=$ diag $(\eta_{\overline{M}\overline{N}},
-\eta_{\underline{M}\underline{N}})$. 
The vielbein field $E_M{}^A$ is O($D,D$) gauge field 
\bea
E_M{}^A\eta^{MN}E_N{}^B=\eta^{AB}~~~.\label{ODDeta}
\eea
With use of the worldsheet doubled zweibeins,
 $\bar{e}_a{}^m$ and  $\underline{e}_a{}^m$,
the Lagrangian \bref{LRgs} is rewritten as
\bea
&L~=~
\displaystyle\frac{1}{\bar{e}}\bar{J}_+{}^{\overline{A}}\eta_{\overline{A}\overline{B}} \bar{J}_-{}^{\overline{B}}
+\frac{1}{\underline{e}} \underline{J}_+{}^{\underline{A}}\eta_{\underline{A}\underline{B}} \underline{J}_-{}^{\underline{B}}\label{gsLxy}&\\
&\left\{\begin{array}{ccl}
\bar{J}_a{}^{\overline{A}}&=&\bar{e}_a{}^m\partial_m X^{{M}}E_M{}^{\overline{A}}\\
\underline{J}_a{}^{\underline{A}}&=&\underline{e}_a{}^m\partial_m X^{{M}}E_M{}^{\underline{A}}
\end{array}\right.&\nn\\
&
e_a{}^m=\left(\begin{array}{cc}
e_-{}^\tau&e_-{}^\sigma\\
e_+{}^\tau&e_+{}^\sigma
\end{array}	\right)~~,~~
\bar{e}_a{}^m=\left(\begin{array}{cc}
1&-(g+s)\\1&\tilde{g}+\tilde{s}\end{array}	\right)~~,~~\label{zweiT}
\underline{e}_a{}^m=\left(\begin{array}{cc}
	1&-(g-s)\\1&\tilde{g}-\tilde{s}\end{array}	\right)~&\nn
\eea
with $\bar{e}=\rm{det}~\bar{e}_a{}^m$ and $\underline{e}=\rm{det}~\underline{e}_a{}^m$.

The Lagrangian 
with the usual single worldvolume zweibein  $e_a{}^m$ 
is given by
\bea
L&=&
\frac{1}{e}\left[
J_+{}^{\overline{A}}\eta_{\overline{A}\overline{B}}  J_-{}^{\overline{B}}
+ J_+{}^{\underline{A}}\eta_{\underline{A}\underline{B}}  J_-{}_{\underline{B}}
-\lambda_+ J_-{}^{\overline{A}}\eta_{\overline{A}\overline{B}}  J_-{}^{\overline{B}}
-\lambda_-  J_+{}^{\underline{A}}\eta_{\underline{A}\underline{B}}  J_+{}_{\underline{B}}
\right]\label{eJJ} \\
&&\quad\quad\quad\quad\quad\quad\quad\quad\left\{\begin{array}{ccl}
	J_a{}^{\overline{A}}&=&{e}_a{}^m\partial_m X^ME_M{}^{\overline{A}}\\
	J_a{}^{\underline{A}}&=&{e}_a{}^m\partial_m X^ME_M{}^{\underline{A}}
\end{array}\right.\nn~\\
&&\quad\quad\quad\quad\quad\quad\quad\quad{e}_a{}^m=\left(\begin{array}{cc}
	1&-(g+s)\\1&{g}-{s}\end{array}	\right) \nn\\
&&\quad\quad\quad\quad\quad\quad\quad\quad
\lambda_\pm=\varphi\{
-(s+\tilde{s}\mp g)^2+\tilde{g}^2\}\nn
\eea
with ${e}=\rm{det}~{e}_a{}^m=2g$ .

It is useful to give the T-theory Lagrangian 
in terms of  
the selfdual and the anti-selfdual currents  
 where the anti-selfdual current is the selfduality constraint.
\bea
L&=&\frac{1}{g}\don\circ{J}_{\rm{SD}}{}^M
G_{MN}\don\circ{J}_{\overline{\rm{SD}}}{}^N
-\hat{\lambda}\don\circ{J}_{\overline{\rm{SD}}}{}^M G_{MN}\don\circ{J}_{\overline{\rm{SD}}}{}^N
-\lambda  
\don\circ{J}_{\overline{\rm{SD}}}{}^M{\eta}_{MN}\don\circ{J}_{\overline{\rm{SD}}}{}^N\label{SDASDTaction}\\
&=&\frac{1}{g}J_{\rm{SD}}{}^A\hat{\eta}_{AB}J_{\overline{\rm{SD}}}{}^B
-\hat{\lambda}J_{\overline{\rm{SD}}}{}^A\hat{\eta}_{AB}J_{\overline{\rm{SD}}}{}^B
-\lambda  J_{\overline{\rm{SD}}}^A{\eta}_{AB}J_{\overline{\rm{SD}}}^B\nn
\\
&&\left\{\begin{array}{ccl}
	J_{\rm{SD}}{}^A&=&E_M{}^A\don\circ{J}_{\rm{SD}}{}^M\\
	J_{\overline{\rm{SD}}}^{}A&=&E_M{}^A\don\circ{J}_{\overline{\rm{SD}}}{}^M
\end{array}\right.\nn\\
&&\left\{\begin{array}{ccl}
	\don\circ{J}_{\rm{SD}}{}^M&=&\dot{X}^M+(g\hat{\eta}-s\eta)^{MN}\eta_{NL}\partial_\sigma X^L\\
	\don\circ{J}_{\overline{\rm{SD}}}{}^M&=&\dot{X}^M-(g\hat{\eta}+s\eta)^{MN}\eta_{NL}\partial_\sigma X^L
\end{array}\right.\nn\\
&&\left\{\begin{array}{ccl}
\hat{\lambda}&=&\displaystyle\frac{1}{g}-
\varphi{(g+\tilde{g})}
\\
\lambda&=&
\varphi({s+\tilde{s}})
\end{array}\right.\nn
\eea
Bilinears of the anti-selfdual currents 
relax the selfduality constraint $\don\circ{J}_{\overline{SD}}{}^M=0$
as shown in \bref{ASDcon}.
It is the selfduality constraint in curved backgrounds
$g^{\tau m}\partial_m X^NG_{NM}=\epsilon^{\tau m}\partial_m X^N\eta_{NM}$ with
$g^{\tau\tau}=\frac{1}{g}$, $g^{\tau\sigma}=-\frac{s}{g}$ and $G_{NM}=\hat{\eta}_{NM}$.
In this formulation the worldsheet zweibein is not factored out in this Lagrangian.

\par
\vskip 6mm
\subsection{SO(5,5) Hamiltonian form action }

We apply the double zweibein formulation to
construct F10-brane actions.
The Hamiltonian is sum of the set of Virasoro constraints  ${\cal T}={\cal S}^{\underline{m}}={\cal U}^\mu=0$ 
 in \bref{Virasoro} 
and Virasoro constraints for the anti-selfdual currents 
$\tilde{\cal T}=\tilde{\cal S}^{\underline{m}}=0$ 
 in \bref{33}
with Lagrange multipliers 
$g,~ s_{\underline{m}},~ Y_\mu~$, $\tilde{g},~ \tilde{s}_{\underline{m}} $ respectively.
$Y_\mu$ plays the role of $A_0$ in the usual gauge theory.
We begin by the following Hamiltonian form action  in the SO(5,5) spinor representation
\bea
I&=&\displaystyle \int d\tau d^{10}\sigma ~L~~,~~
L~=~\dot{X}^\mu P_\mu-H\nn\\
H&=&\left(g{\cal T}+s_{\underline{m}} {\cal S}^{\underline{m}}
\right)+
\left(\tilde{g} \tilde{\cal T}+\tilde{s}_{\underline{m}} \tilde{\cal S}^{\underline{m}}\right)
+{\cal U}^\mu Y_\mu\label{actionSO55}\\
&=&\frac{1}{4}\dd_\mu
(g\hat{\eta}+s_{\underline{m}}  \gamma^{\underline{m}} )^{\mu\nu}
\dd_\nu
+\frac{1}{4}\tilde{\dd}_\mu(\tilde{g}\hat{\eta}+\tilde{s}_{\underline{m}}\gamma^{\underline{m}})^{\mu\nu}\tilde{\dd}_\nu
+\dd_\mu (\slashed{\partial} Y)^\mu ~~~.\nn
\eea
The Lagrangian is written in terms of field strengths $\don\circ{F}_\pm{}^\mu$ as 
\bea
&L=\varphi
\don\circ{F}_+{}^\mu
\left\{(g+\tilde{g})\hat{\eta}-(s+\tilde{s})_{\underline{m}} \gamma^{\underline{m}}   )\right\}_{\mu\nu}
\don\circ{F}_-{}^\nu
+Y_\mu {\cal V} X^\mu&
~~~\label{SO55Hamaction}\\
&
\left\{\begin{array}{ccl}
\don\circ{F}_+{}^\mu&=&\don\circ{F}_{\tau}{}^\mu
+(\tilde{g}\hat{\eta}+\tilde{s}^{\underline{m}}\gamma_{\underline{m}})^{\mu\nu}
\don\circ{F}_{\sigma~\nu}\\
\don\circ{F}_-{}^\mu&=&\don\circ{F}_{\tau}{}^\mu
-({g}\hat{\eta}+{s}^{\underline{m}}\gamma_{\underline{m}})^{\mu\nu}
\don\circ{F}_{\sigma~\nu}
\end{array}\right.~~,~~
\left\{\begin{array}{ccl}
\don\circ{F}_{\tau}{}^\mu&=&\dot{X}^\mu-\slashed{\partial}^{\mu\nu} Y_\nu\\
\don\circ{F}_{\sigma~\mu}&=&\slashed{\partial}_{\mu\nu} X^{\nu}\end{array}\right.
\label{FF}&~~~\nn\\
&\varphi=
\left[{(g+\tilde{g})^2-(s+\tilde{s})^{\underline{m}}
	(s+\tilde{s})_{\underline{m}}}\right]^{-1}&\nn
\eea
and the ${\cal V}=0$ constraint given in \bref{Virasoro}.
The field strengths are invariant under 
the gauge transformation with the gauge parameter $\kappa_\mu$ and $\bar{\kappa}^\mu$
\bea
&\delta_\kappa X^\mu=\slashed{\partial}^{\mu\nu}\kappa_\nu~~,~~
\delta_\kappa Y_\mu=\dot{\kappa}_{\mu}+ \slashed{\partial}_{\mu\nu}\bar{\kappa}^\nu &
~~~
\eea
by using ${\cal V}=0$ constraint.
This gauge transformation is generated by the Gauss law constraint as $\delta_\kappa X^\mu=[\int d\sigma~\kappa_\nu~ {\cal U}^\nu,X^\mu]$.
There are gauge symmetries of the gauge symmetry 
as same as the $\kappa$-symmetry
$\delta \kappa=\slashed{\partial} \kappa^{[1]}$, $\delta\kappa^{[1]}=\slashed{\partial} \kappa^{[2]}$, $\cdots$ and 
$\delta \bar{\kappa}=\slashed{\partial}\bar{\kappa}_{[1]}$, 
$\delta \bar{\kappa}_{[1]}= \slashed{\partial}\bar{\kappa}_{[2]}$, $\cdots$.
The infinite series of gauge symmetries reduce a half of the coordinates.
The 10-dimensional covariant action requires the auxiliary coordinate $Y_\mu$, but it is removed by the bosonic $\kappa$ symmetry in the temporal gauge.
In the lightcone-like gauge a half of $X^\mu$ is removed, giving the (4+4)-dimensional T-theory.

	In  F-theory currents $F_\pm{}^\mu$  
the worldsheet vielbein  cannot be extracted 
from the spacetime vielbein
because the SO(5,5) covariant $\gamma$-matrix 
$\gamma_{\underline{m}\mu\nu}$ mixes the worldvolume index and the spacetime index
unlikely to the O($D,D$) invariant metric $\eta_{MN}$ in the T-theory in \bref{ODDeta}.
The first term contains both the selfdual and the anti-selfdual field strengths
so it is a free kinetic term.
Other terms contain only the anti-selfdual field strength, then they  
are constraints.
The F10-brane Lagrangian is further rewritten in terms of the selfdual and the anti-selfdual currents analogously to \bref{SDASDTaction} as
\bea
L&=&\frac{1}{g}
\don\circ{F}_{\rm{SD}}{}^\mu\hat{\eta}_{\mu\nu}
\don\circ{F}_{\overline{\rm{SD}}}{}^\nu
-\hat{\lambda}
\don\circ{F}_{\overline{\rm{SD}}}{}^\mu\hat{\eta}_{\mu\nu}
\don\circ{F}_{\overline{\rm{SD}}}{}^\nu
-\lambda^{\underline{m}}
\don\circ{F}_{\overline{\rm{SD}}}{}^\mu{\gamma}_{\underline{m}\mu\nu}
\don\circ{F}_{\overline{\rm{SD}}}{}^\nu
\label{SDASDFaction}\\
&&\left\{\begin{array}{ccl}
	\don\circ{F}_{\rm{SD}}{}^\mu&=&\don\circ{F}_{\tau}{}^\mu
	+(g\hat{\eta}-s_{\underline{m}}
	\gamma^{\underline{m}})^{\mu\nu}\don\circ{F}_{\sigma~\nu} \\
	\don\circ{F}_{\overline{\rm{SD}}}{}^\mu&=&\don\circ{F}_{\tau}{}^\mu
	-(g\hat{\eta}+s_{\underline{m}}\gamma^{\underline{m}})^{\mu\nu}\don\circ{F}_{\sigma~\nu}
\end{array}\right.\nn\\
&&
\left\{\begin{array}{ccl}
	\hat{\lambda}&=&\displaystyle\frac{1}{g}-\varphi
	({g+\tilde{g}})
	\nn\\
	\lambda_{\underline{m}}&=&
	\varphi(s+\tilde{s})_{\underline{m}}
\end{array}\right.\label{FSDASDL}
\eea
The selfdual and the anti-selfdual currents in F-theory 
mixes the worldsheet vielbein and the spacetime vielbein
as in  \bref{SDASDTaction}.

In curved background 
the SO(5,5) gauge fields $G_{\mu\nu}$ and the SO(5,5) currents 
are given with the SO(5,5)/SO(5;$\mathbb{C}$) vielbein  $E_A{}^{M}$ as
\bea
&G_{\mu\nu}=E_\mu{}^\alpha 
\hat{\eta}_{\alpha\beta}E_\nu{}^\beta~~,~~
F_{\pm}{}^\alpha=E_\mu{}^\alpha\don\circ{F}_\pm{}^\mu\label{Efs}&~~~.
\eea
The F10-brane Lagrangian in curved background 
is written  as
\bea
L&=&\varphi
\don\circ{F}_+{}^\mu
\left\{(g+\tilde{g})G_{\mu\nu}-(s+\tilde{s}){}^{\underline{m}} \gamma_{\underline{m}\mu\nu}   )\right\}
\don\circ{F}_-{}^\nu
\label{Fcurvepmaction}\\
&=&\varphi
{F}_+{}^\alpha
\left\{(g+\tilde{g})\hat{\eta}_{\alpha\beta}
-(s+\tilde{s}){}^{\underline{a}} \gamma_{\underline{a}\alpha\beta}   )\right\}
{F}_-{}^\beta
\nn~~~
\eea
with  SO(5,5) vector parameter 
 $(s+\tilde{s})^{\underline{a}}=(s+\tilde{s})^{\underline{m}}
 {\cal E}_{\underline{m}}{}^{\underline{a}}$, as
 \bref{SO55gamma}.

Then now we propose a Lagrangian for a F10-brane in curved background in terms of the selfdual and the anti-selfdual currents is given as
\bea
L&=&\frac{1}{g}
\don\circ{F}_{\rm{SD}}{}^\mu 
G_{\mu\nu}
\don\circ{F}_{\overline{\rm{SD}}}{}^\nu
-\hat{\lambda}
\don\circ{F}_{\overline{\rm{SD}}}{}^\mu G_{\mu\nu}
\don\circ{F}_{\overline{\rm{SD}}}{}^\nu
-\lambda^{\underline{m}}
\don\circ{F}_{\overline{\rm{SD}}}{}^\mu{\gamma}_{\underline{m}\mu\nu}
\don\circ{F}_{\overline{\rm{SD}}}{}^\nu
\label{FSDaction}\\
&=&\frac{1}{g}
{F}_{\rm{SD}}{}^\alpha 
\hat{\eta}_{\alpha\beta}
{F}_{\overline{\rm{SD}}}{}^\beta
-\hat{\lambda}
{F}_{\overline{\rm{SD}}}{}^\alpha \hat{\eta}_{\alpha\beta}
{F}_{\overline{\rm{SD}}}{}^\beta
-\lambda^{\underline{a}}
{F}_{\overline{\rm{SD}}}{}^\alpha{\gamma}_{\underline{a}\alpha\beta}
{F}_{\overline{\rm{SD}}}{}^\beta
\nn\\
&&\left\{\begin{array}{ccl}
F_{\rm{SD}}{}^\alpha&=&E_\mu{}^\alpha\don\circ{F}_{\rm{SD}}{}^\mu\\
F_{\overline{\rm{SD}}}{}^\alpha&=&E_\mu{}^\alpha\don\circ{F}_{\overline{\rm{SD}}}{}^\mu
\end{array}\right.\nn
\eea
The $\hat{\lambda}$ and $\lambda^{\underline{a}}$ 
are Lagrange multipliers for selfduality constraints  given in 
\bref{SDASDFaction}.

\par
\vskip 6mm
\subsection{SO(6,6) Lagrangian form action}

Next let us consider the SO(6,6) F-symmetry covariant action.
The SO(6,6) $\gamma$-matrix is given by 64$\times$64 matrix $\Gamma^{\underline{\hat{a}}}$ 
with two 32$\times$32 matrices $\Sigma^{\underline{\hat{a}}}$ and $\tilde{\Sigma}^{\underline{\hat{a}}}$ as
\bea
&\Gamma^{\underline{\hat{a}}}=
\left(\begin{array}{cc} 0&
\tilde{\Sigma}^{\underline{\hat{a}}}{}^{\underline{\alpha'}
	\underline{\beta}}\\
{\Sigma}^{\underline{\hat{a}}}{}_{\underline{\alpha}\underline{\beta'}}&0\end{array}
\right)\label{Gamma12}~~,~~\underline{\hat{a}}=1,\cdots,12~~~.
\eea
They satisfy the following algebra with the SO(6,6) invairant metric $\eta^{\underline{\hat{a}\hat{b}}}$.
\bea
\{\Gamma^{\underline{\hat{a}}},
\Gamma^{\underline{\hat{b}}}\}
&=&2\eta^{\underline{\hat{a}\hat{b}}}~~,~~
{\renewcommand{\arraystretch}{1.6}\left\{\begin{array}{l}
\Sigma^{(\underline{\hat{a}}}{}_{\underline{\alpha}\underline{\beta}'}
\bar{\Sigma}^{\underline{\hat{b}})\underline{\beta}'\underline{\gamma}}=2\eta^{\underline{\hat{a}}\underline{\hat{b}}}\delta_{\underline{\alpha}}^{\underline{\gamma}}\\
\bar{\Sigma}^{(\underline{\hat{a}}}{}^{\underline{\alpha}'\underline{\beta}}
{\Sigma}^{\underline{\hat{b}})}{}_{\underline{\beta}\underline{\gamma}'}
=2\eta^{\underline{\hat{a}}\underline{\hat{b}}}\delta^{\underline{\alpha}'}_{\underline{\gamma}'}
\end{array}\right.}
\eea
The SO(5,5) $\gamma$-matrix is embedded with $\underline{\hat{a}}=(+,-,\underline{a})$, $\underline{a}=(1,\cdots,10)$ as
\bea
\Sigma^{\underline{\hat{a}}}{}_{\underline{\alpha}\underline{\beta}'}&:&
\Sigma^+=\left(\begin{array}{cc}\delta_\alpha^\beta&0\\0&0\end{array}\right)~~,~~
\Sigma^-=\left(\begin{array}{cc}0&0\\0&\delta_\beta^\alpha\end{array}\right)~~,~~
\Sigma^{\underline{a}}=\left(\begin{array}{cc}
0	&\gamma^{\underline{a}}{}_{\alpha\beta}
	\\\gamma^{\underline{a}}{}^{\alpha\beta}&0
\end{array}\right)\nn\\
\tilde{\Sigma}^{\underline{\hat{a}}}{}^{\underline{\alpha}'\underline{\beta}}&:&
\tilde{\Sigma}^+=\left(\begin{array}{cc}0&0\\0&\delta_\alpha^\beta\end{array}\right)~~,~~
\tilde{\Sigma}^-=\left(\begin{array}{cc}\delta_\beta^\alpha&0\\0&0\end{array}\right)~~,~~
\tilde{\Sigma}^{\underline{a}}=-
\left(\begin{array}{cc}
0	&\gamma^{\underline{a}}{}_{\alpha\beta}
   \\\gamma^{\underline{a}}{}^{\alpha\beta}&0
\end{array}\right)
\nn\\
C
_{\underline{\alpha}\underline{\beta}}&=&
\left(\begin{array}{cc}0&-\delta_\alpha^\beta
	\\\delta^\alpha_\beta&0
\end{array}\right)~~,~~
C
_{\underline{\alpha'}\underline{\beta'}}=
\left(\begin{array}{cc}0&-\delta^\alpha_\beta
	\\\delta_\alpha^\beta&0
\end{array}\right)
\label{Rep12gGamma}
\eea
where $\gamma^{\underline{a}}$ is the 10-dimensional $\gamma$-matrix 
$\gamma^{(\underline{a}}{}_{\alpha\beta}\gamma^{\underline{b})}{}^{\beta\gamma}=2\eta^{\underline{ab}}\delta_\alpha^\gamma$ with
$\gamma^{\underline{a}}{}^{\alpha\beta}
=\gamma^{\underline{a}}{}^{\beta\alpha}$.
The SO(6,6) generators are decomposed into the SO(5,5)
dilatation, transformation and rotation  as
\bea
&&\Gamma^{\underline{\hat{a}\hat{b}}}{}_{\underline{\alpha}}{}^{\underline{\beta}}
=\frac{1}{2}(\Sigma^{[\underline{\hat{a}}}\tilde{\Sigma}^{\underline{\hat{b}}]})
{}_{\underline{\alpha}}{}^{\underline{\beta}}
\equiv \Sigma^{\underline{\hat{a}\hat{b}}}{}_{\underline{\alpha}}{}^{\underline{\beta}}
\\
&&\Sigma^{+-}{}_{\underline{\alpha}}{}^{\underline{\beta}}
=\left(\begin{array}{cc}\delta_{\alpha}{}^{\beta}&0\\0&-\delta^{\alpha}{}_{\beta}\end{array}\right)
~,~
\Sigma^{\underline{ab}}{}_{\underline{\alpha}}{}^{\underline{\beta}}
=-\left(\begin{array}{cc}\gamma^{\underline{ab}}{}_\alpha{}^\beta&0\\
0	&\gamma^{\underline{ab}}{}^\alpha{}_\beta\end{array}\right)
\nn\\
&&
\Sigma^{-\underline{a}}{}_{\underline{\alpha}}{}^{\underline{\beta}}
=\left(\begin{array}{cc}0&0\\-\gamma^{\underline{a}}{}^{\alpha\beta}&0\end{array}\right)~,~
\Sigma^{+\underline{a}}{}_{\underline{\alpha}}{}^{\underline{\beta}}
=\left(\begin{array}{cc}0&-\gamma^{\underline{a}}{}_{\alpha\beta}\\0&0\end{array}\right)~~~.
\nn
\eea
We use $\Sigma$'s as the  12-dimensional Weyl-spinors.
It is also convenient to have explicit notation of matrices 
\bea
&&(C\Gamma^{\underline{\hat{a}\hat{b}}})^{\underline{\alpha}\underline{\beta}}
=\frac{1}{2}(C\Sigma^{[\underline{\hat{a}}}\tilde{\Sigma}^{\underline{\hat{b}}]})^{\underline{\alpha}\underline{\beta}}\label{12dSigma}
\\
&&(C\Sigma^{+-})^{\underline{\alpha}\underline{\beta}}
=\left(\begin{array}{cc}0&-\delta^{\alpha}_{\beta}\\-\delta_{\alpha}^{\beta}&0\end{array}\right)
~,~
(C\Sigma^{\underline{ab}})^{\underline{\alpha}\underline{\beta}}
=\left(\begin{array}{cc}0&-\gamma^{\underline{ab}}{}^\alpha{}_\beta\\
\gamma^{\underline{ab}}{}_\alpha{}^\beta&0\end{array}\right)
\nn\\
&&
(C\Sigma^{-\underline{a}})^{\underline{\alpha}\underline{\beta}}
=\left(\begin{array}{cc}-\gamma^{\underline{a}}{}^{\alpha\beta}&0\\0&0\end{array}\right)~,~
(C\Sigma^{+\underline{a}})^{\underline{\alpha}\underline{\beta}}
=\left(\begin{array}{cc}0&0\\0&\gamma^{\underline{a}}{}_{\alpha\beta}\end{array}\right)~~~.
\nn
\eea

Let's rewrite the SO(5,5) covariant action 
\bref{FSDaction} in a SO(6,6) covariant way.
Two 16-component Majorana-Weyl 10-dimensional spinors 
$X^\mu$ and $Y_\mu$ are embedded into a 32-component Majorana-Weyl  12-dimensional spinor as
\bea
Z^{\underline{\alpha'}}=
\left(\begin{array}{c}-Y_\alpha\\X^\alpha\end{array}\right)
~~,~~\underline{\alpha}'=1,\cdots,32~~.
\eea
 The SO(6,6) covariant field strength in a flat background is given by
\bea
\don\circ{F}_{\underline{\alpha}}&=&
\Sigma^{\underline{\hat{a}}}_{\underline{\alpha}\underline{\beta}'}
\partial_{\underline{\hat{a}}}Z^{\underline{\beta}'}
=\left(\begin{array}{c}F_{\sigma~\alpha}\\
F_{\tau}{}^\alpha
\end{array}\right)~~,~~\underline{\alpha}=1,\cdots,32
\nn\\
&=&
\left(\begin{array}{cc}
\delta_\alpha^\beta\partial^-&\slashed{\partial}_{\alpha\beta}\\
\slashed{\partial}^{\alpha\beta}&\delta^\alpha_\beta\partial^+
\end{array}\right)
\left(\begin{array}{c}-Y_\beta\\
X^\beta
\end{array}\right)=\left(\begin{array}{c}
(\slashed{\partial} X)_\alpha-\partial^- Y_\alpha\\
\partial^+ X^\alpha-(\slashed{\partial} Y)^\alpha
\end{array}\right)~~~.\label{dGammaZ}
\eea
The field strengths are invariant under the gauge symmetry 
\bea
\delta_\kappa Z^{\underline{\alpha}'}&=&
\Sigma^{\underline{\hat{a}}}{}^{\underline{\alpha}'\underline{\beta}}
\partial_{\underline{\hat{a}}}\kappa_{\underline{\beta}}\label{boskapS066}~~~.
\eea
The ${\cal V}=0$ constraint in \bref{Virasoro}
is enlarged SO(6,6) covariantly.
\bea
&&{\cal V}
=\eta_{\underline{mn}}\partial^{\underline{m}}\partial^{\underline{n}}=0~~\rm{with}~~\partial^-=0~~\nn\\
&&\to~~
\hat{\cal V}=\don\circ{\eta}_{\underline{\hat{m}\hat{n}}} \partial^{\underline{\hat{m}}}\partial^{\underline{\hat{n}}}=
\eta_{\underline{mn}}\partial^{\underline{m}}\partial^{\underline{n}}
+2\partial^{+}\partial^{-}
=0
\eea

We propose a F10-brane Lagrangian in a flat space with the worldvolume vielbein
by rewriting the SO(5,5) covariant action 
\bref{FSDaction}
with the SO(6,6) field strength \bref{dGammaZ} with the 
32$\times$32 matrix metric $\don\circ{\eta}^{\underline{\alpha}\underline{\beta}}$ as
\bea
L&=&
e{F}_{\underline{\alpha}}
\don\circ{\eta}^{\underline{\alpha}\underline{\beta}}
{F}_{\underline{\beta}}
-\frac{1}{2}\lambda_{\underline{\hat{a}\hat{b}}} 
{F}_{\underline{\alpha}}
(C\Sigma^{\underline{\hat{a}\hat{b}}}) ^{\underline{\alpha}\underline{\beta}}
{F}_{\underline{\alpha}}\nn\\
&=&
e\don\circ{F}_{\underline{\alpha}}(E^T
\don\circ{\eta}E
)^{\underline{\alpha}\underline{\beta}}
\don\circ{F}_{\underline{\beta}}
-\frac{1}{2}\lambda_{\underline{\hat{a}\hat{b}}} 
\don\circ{F}_{\underline{\alpha}}
(E^T
C\Sigma^{\underline{\hat{a}\hat{b}}}E%
) ^{\underline{\alpha}\underline{\beta}}
\don\circ{F}_{\underline{\alpha}}
~~~\label{SO66action}\\
&&\don\circ{\eta}^{\underline{\alpha}\underline{\beta}}=\left(\begin{array}{cc}
	-\hat{\eta}^{\alpha\beta}&0\\0&\hat{\eta}_{\alpha\beta}\end{array}\right)~~~.\nn
\eea
The field strength with the worldvolume vielbein  in a flat space
$F_{\underline{\alpha}}$ is given by
\bea
F_{\underline{\alpha}}&=&\rm{exp}
(\frac{1}{2} f_{\underline{ab}} \Sigma^{\underline{ab} } )
{}
_{\underline{\alpha}}{}^{\underline{\beta}}
\don\circ{F}_{\underline{\beta}}\nn\\
\rm{exp}
(\frac{1}{2} f_{\underline{ab}} \Sigma^{\underline{ab} } )
{}_{\underline{\alpha}}{}^{\underline{\beta}}&=&
\rm{exp}(\frac{1}{2}\rm{ln}\phi~\Sigma^{+-})_{\underline{\alpha}}{}^{\underline{\gamma}}~~
\rm{exp}(-\psi_{\underline{a}}~\Sigma^{-\underline{a}})_{\underline{\gamma}}{}^{\underline{\beta}}~~~\label{Ewv}\\
&=&\left(\begin{array}{cc}
\sqrt{\phi}~\delta_\alpha^\gamma &0\\
0&\displaystyle\frac{1}{\sqrt{\phi}}~\delta_\gamma^\alpha \end{array}\right)
\left(\begin{array}{cc}
	\delta_\gamma^\beta&0	\\
	\psi_{\underline{a}}\gamma^{\underline{a}\gamma\beta}	
	&\delta_\beta^\gamma\end{array}\right)\nn
\eea
with $f_{+-}=\rm{ln}\phi$ and $f_{-\underline{a}}=-\psi_{\underline{a}}$.
The Gauss law constraint is derived from the Lagrangian analogously to the usual gauge theory action.
Lagrange multipliers, 
 $e$,~$\lambda_{\underline{\hat{a}\hat{b}}}$, 
and the worldvolume vielbein fields $\phi,~\psi_{\underline{a}}$, in the SO(6,6) action \bref{SO66action} 
correspond to Lagrange multipliers of Virasoro constraints
for selfdual and  anti-selfdual parts, 
$g$, $s_{\underline{a}}$ and $\tilde{g}$,  $\tilde{s}_{\underline{a}}$,
 in the SO(5,5) action \bref{Fcurvepmaction} as
\bea
&&
\left\{{\renewcommand{\arraystretch}{1.8}
		\begin{array}{ccl}
e&=&\sqrt{\varphi g\tilde{g}}\\
\phi&=&\sqrt{\displaystyle\frac{g\tilde{g}}{\varphi}}\displaystyle\frac{1}{g+\tilde{g}}\\
\psi_{\underline{a}}&=&\displaystyle\frac{g\tilde{s}_{\underline{a}}-\tilde{g}s_{\underline{a}}}{g+\tilde{g}}
\end{array}}\right.
~~,~~
\left\{{\renewcommand{\arraystretch}{1.8}
	\begin{array}{ccl}
			\lambda_{+-}&=&\displaystyle\frac{g-\tilde{g}}{2(g+\tilde{g})}\\
		\lambda_{\underline{ab}}&=&
		\varphi \tilde{s}_{[\underline{a}}{s}_{\underline{b}]}\\
		\lambda_{+\underline{a}}&=&-\displaystyle\frac{\sqrt{g\tilde{g}\varphi}}{g+\tilde{g}}
		(s_{\underline{a}}+\tilde{s}_{\underline{a}})\\
		\lambda_{-\underline{a}}&=&
		-\lambda_{+\underline{a}}-\displaystyle\frac{2}{\phi}\lambda_{\underline{ab}}\psi^{\underline{b}}
	\end{array}}\right.
\label{lambdags}%
\eea
with $\varphi^{-1}={(g+\tilde{g})^2-(s+\tilde{s})^2}$.
They are solved inversely as
\bea
&&
\left\{{\renewcommand{\arraystretch}{1.8}
	\begin{array}{ccl}
g&=&\displaystyle\frac{2e\phi}{1-2\lambda_{+-}}\\
\tilde{g}&=&\displaystyle\frac{2e\phi}{1+2\lambda_{+-}}\\
s_{\underline{a}}&=&
-\psi_{\underline{a}}-\displaystyle\frac{2\phi}{1-2\lambda_{+-}}\lambda_{+\underline{a}}\\
\tilde{s}_{\underline{a}}&=&
\psi_{\underline{a}}-\displaystyle\frac{2\phi}{1+2\lambda_{+-}}\lambda_{+\underline{a}}
\end{array}}\right.
\label{lambdags2}
\eea

We finally obtain the SO(6,6) covariant action for the F10-brane  in  curved backgrounds 
\bea
I&=&\int d^{12}\sigma ~L\nn\\
L&=&e\don\circ{F}{}_{\underline{\mu}}
G^{\underline{\mu\nu}} \don\circ{F}{}_{\underline{\nu}}
-\frac{1}{2}\lambda_{\underline{\hat{m}\hat{n}}}\don\circ{F}{}_{\underline{\mu}}
(C\Sigma^{\underline{\hat{m}\hat{n}}})^{\underline{\mu\nu}} \don\circ{F}{}_{\underline{\nu}}\label{FSO66}\\%
&=& {F}{}_{\underline{\alpha}}
\don\circ{\eta}^{\underline{\alpha\beta}} {F}{}_{\underline{\beta}}
-\frac{1}{2}\lambda_{\underline{\hat{a}\hat{b}}}
{F}{}_{\underline{\alpha}}
(C\Sigma^{\underline{\hat{a}\hat{b}}})^{\underline{\alpha\beta}} {F}{}_{\underline{\beta}}
\nn\\
G^{\underline{\mu\nu}}&=&
E_{\underline{\alpha}}{}^{\underline{\mu}}
\don\circ{\eta}^{\underline{\alpha \beta}}
E_{\underline{\beta}}{}^{\underline{\nu}}~~
\nn
\eea
where $\lambda_{\underline{\hat{m}\hat{n}}}$ and $\lambda_{\underline{\hat{a}\hat{b}}}$ are related by the worldvolume vielbein  as \bref{lambdags}.
The SO(6,6) covariant field strength in a curved background $F_{\underline{\alpha}}$ 
and the one in a flat space $\don\circ{F}_{\underline{\mu}}$ are given as
\bea
{F}{}_{\underline{\alpha}}&=&E_{\underline{\alpha}}{}^{\underline{{\mu}}}\don\circ{F}{}_{\underline{\mu}}=(\Sigma^{\underline{\hat{a}}})_{\underline{\alpha\beta}'}
{\cal E}_{\underline{\hat{a}}}{}^{\underline{\hat{m}}}
\partial_{\underline{\hat{m}}}Z^{\underline{\mu}'}
E_{\underline{\mu}'}{}^{\underline{\beta}'}\\
\don\circ{F}{}_{\underline{\mu}}&=&
(\Sigma^{\underline{\hat{m}}})_{\underline{\mu\nu}'} \partial_{\underline{\hat{m}}}Z^{\underline{\nu}'}\nn
\eea
The SO(6,6) vielbein satisfies the following condition
where the SO(5,5) vielbein field is embedded in
the SO(6,6) vielbein field as
\bea
\Sigma^{\underline{\hat{a}}}{}_{\underline{\alpha\beta}'}&=&
\Sigma^{\underline{\hat{m}}}{}_{\underline{\mu\nu}'}
E_{\underline{\alpha}}{}^{\underline{\mu}}
E_{\underline{\beta}'}{}^{\underline{\nu}'}
{\cal E}_{\underline{\hat{m}}}{}^{\underline{\hat{a}}}~~~\label{orthogonalcon}
\\
E_{\underline{\alpha}}{}^{\underline{\mu}}&=&\rm{exp}
(\frac{1}{2} f_{\underline{ab}} \Sigma^{\underline{ab} } )
{}
_{\underline{\alpha}}{}^{\underline{\beta}}
\left(\begin{array}{cc}
E_\beta{}^\mu&0\\0&(E_{\mu}{}^{\beta})^T
\end{array}\right)
\nn\\
E_{\underline{\alpha}'}{}^{\underline{\mu}'}&=&
\left(\begin{array}{cc}
	(E_{\mu}{}^{\beta})^T&0\\0&E_{\beta}{}^{\mu}
\end{array}\right)
\rm{exp}
(\frac{1}{2} f_{\underline{ab}} \Sigma^{\underline{ab} } )
{}^{\underline{\beta}'}{}_{\underline{\alpha}'}~~~.\nn
\label{stSO55}
\eea

Both the SO(5,5) spacetime vielbein  and the worldvolume vielbein combine into the vielbein field of the SO(6,6) F-theory.
The SO(6,6) F-theory background  
is described by the coset SO(6,6)/SO(6;$\mathbb{C}$) 
with its dimension $36=25+11$.
The number of spacetime vielbein, SO(5,5)/SO(5;$\mathbb{C}$) fields  is 25, 
while the number of Virasoro constraints of a F10-brane is 11.
The SO(6,6) vielbein is transformed under the SO(6,6) transformation as
\bea
&\Lambda_{\underline{\hat{a}}\underline{\hat{b}}}\in~\rm{SO}(6,6)~~,~~
E_{\underline{\hat{\mu}}}{}^{\underline{\hat{\alpha}}}
\to E_{\underline{\hat{\mu}}}{}^{\underline{\hat{\beta}}}~
\rm{exp}\left(\Lambda_{\underline{\hat{a}}\underline{\hat{b}}}
\Gamma^{\underline{\hat{a}}\underline{\hat{b}}}
\right){}_{\underline{\hat{\beta}}}{}^{\underline{\hat{\alpha}}}~~,~~
{\cal E}_{\underline{\hat{a}}}{}^{\underline{\hat{m}}}
\to \Lambda_{\underline{\hat{a}}}{}^{\underline{\hat{b}}}{\cal E}_{\underline{\hat{b}}}{}^{\underline{\hat{m}}}~~~.\nn&
\eea

\par

\vskip 6mm
\subsection{GL(6) and GL(5) actions}

In order to construct the perturbative M-theory 5-brane action
we rewrite the SO(6,6) 
F10-brane action obtained in the previous section \bref{FSO66}
to the ones with GL(6) and GL(5) symmetries.
The SO(6,6) spinor representations are decomposed into the 
GL(6) and GL(5) tensor representations as follows: 
\bea
&\begin{array}{l}
	\rm{F}\mathchar`-\rm{symmetry}~\rm{of}~\rm{F}\mathchar`-\rm{theory}\\	\rm{SO}(6,6)\\
	~{\renewcommand{\arraystretch}{1.2}\left\{\begin{array}{l}
		Z^{\underline{\mu}}(32)\\
		F_{\underline{\mu}}(32)\\
		\partial_{\underline{\hat{m}}}(12)
	\end{array}\right.}
\end{array}&\nn\\
&\quad\quad\quad\swarrow
\quad\quad\quad\quad\quad\quad\quad\quad\quad\searrow\quad\quad\quad\quad
&\nn\\
&\begin{array}{l}
\rm{F}\mathchar`-\rm{symmetry}~\rm{of}~\rm{M}\mathchar`-\rm{theory}\\	\rm{GL}(6)\\~
{\renewcommand{\arraystretch}{1.2}\left\{\begin{array}{rcl}
		Z^{\underline{M}}(32)&=&Z^{\hat{m}}(6)\oplus \bar{Z}_{\hat{m}}(6)\oplus Z_{\hat{m}\hat{n}\hat{l}}(20)\\
		F_{\underline{M}}(32)&=&F_{\hat{m}\hat{n}}(15)\oplus F^{\hat{m}\hat{n}}(15)\oplus  F(1)\oplus\bar{F}(1)\\
		\partial_{\underline{\hat{m}}}(12)&=&{\partial}^{\hat{m}}(6)\oplus \partial_{\hat{m}}(6)
	\end{array}\right.}\end{array}
~
\begin{array}{l}
\rm{G}\mathchar`-\rm{symmetry}~\rm{of}~\rm{F}\mathchar`-\rm{theory}\\\rm{SO}(5,5)\\
	~{\renewcommand{\arraystretch}{1.2}\left\{\begin{array}{l}		X^\mu(16)\oplus Y_\mu(16)\\		F_{\mu}(16)\oplus F^{\mu}(16)	\\		\partial_{{m}}(5)\oplus \bar{\partial}^m(5)\oplus \partial^{+}(1)		\oplus {\partial}^-(1)		\end{array}\right.}
\end{array}&
\nn\\
&\quad\quad\quad\searrow
\quad\quad\quad\quad\quad\quad\quad\quad\quad\swarrow\quad\quad\quad\quad
&\nn\\
&\begin{array}{l}
\rm{G}\mathchar`-\rm{symmetry}~\rm{of}~\rm{M}\mathchar`-\rm{theory}\\	\rm{GL}(5)\\~
	{\renewcommand{\arraystretch}{1.2}	\left\{	\begin{array}{ccl}		
Z^{\hat{m}}(6)&=&X^{{m}}(5)\oplus {Y}(1)\\
\bar{Z}_{\hat{m}}(6)&=&Y_{m}(5)\oplus 	 \bar{X}(1)\\
Z^{\hat{m}\hat{n}\hat{l}}(20)&=&	X_{mn}(10)\oplus  Y^{mn}(10)\\
F_{\hat{m}\hat{n}}(15)&=&F_{\tau~{m}{n}}(10)\oplus  F_{\sigma~m}(5)\\
F^{\hat{m}\hat{n}}(15)&=&F_{\sigma}^{{m}{n}}(10)\oplus  F_{\tau}^{m}(5)\\
F(1)&=&F_{\sigma}(1)\\
\bar{F}(1)&=&\bar{F}_{\tau}\\
\partial^{\hat{m}}(6)&=&\partial^{{m}}(5)\oplus \partial^{+}
(1)\\
\partial_{\hat{m}}(6)&=&\partial_{{m}}(5)\oplus \partial^{-}
(1)	\end{array}\right.}\end{array}\nn
&\nn\\
\eea

The GL(6) and GL(5) covariant field strengths, 
denoted by $F_{\underline{M}}$ instead of  $\don\circ{F}_{\underline{M}}$ for simplicity, and their gauge transformation rules  are given as below. 
\bea
&&\rm{GL(6)~field~strengths}\nn\\
&&\quad\quad {F}{}_{\underline{M}}=(F^{\hat{m}\hat{n}},~
F_{\hat{m}\hat{n}},~F,~\bar{F})~~,~~\hat{m}=1,\cdots,6\nn\\
&&\quad\quad{\renewcommand{\arraystretch}{1.2} \left\{
\begin{array}{ccl}
	F^{\hat{m}_1\hat{m}_2}&=&\partial^{[\hat{m}_1}Z^{\hat{m}_2]}
	+\frac{1}{3!}
	\epsilon^{\hat{m}_1\cdots \hat{m}_6}\partial_{\hat{m}_3}Z_{\hat{m}_4\cdots\hat{m}_6}\\
	F_{\hat{m}_1\hat{m}_2}&=&\partial_{[\hat{m}_1}\bar{Z}_{\hat{m}_2]}
	+\partial^{\hat{l}}Z_{\hat{m}_1\hat{m}_2\hat{l}}\\
	F&=&\partial_{\hat{m}}Z^{\hat{m}}\\
	\bar{F}&=&\partial^{\hat{m}}\bar{Z}_{\hat{m}}
\end{array}
\right.}\label{FSGL6}
\\
\nn\\&&\rm{Gauge~transformations
}\nn\\
&&\quad\quad \kappa_{\underline{M}}=(\kappa^{\hat{m}\hat{n}},~\kappa_{\hat{m}\hat{n}},~\kappa,~\bar{\kappa})
\nn\\&&\quad\quad {\renewcommand{\arraystretch}{1.2}\left\{
\begin{array}{ccl}
	\delta_\kappa Z^{\hat{m}}&=&\partial^{\hat{m}}\kappa
	+\partial_{\hat{n}}\kappa^{\hat{n}\hat{m}}\\
	\delta_\kappa \bar{Z}_{\hat{m}}&=&\partial_{\hat{m}}\bar{\kappa}
	+\partial^{\hat{n}}\kappa_{\hat{n}\hat{m}}\\
	\delta_\kappa Z_{\hat{m}_1\hat{m}_2\hat{m}_3}&=&\frac{1}{2}
	\partial_{[\hat{m}_1}\kappa_{\hat{m}_2\hat{m}_3]}
	+\frac{1}{2}\epsilon_{\hat{m}_1\cdots \hat{m}_6}
	\partial^{\hat{m}_4}\kappa^{\hat{m}_5\hat{m}_6}
\end{array}\right.}\label{Zkappa6}
\eea
\bea
&&\rm{GL(5)~field~strengths}\nn\\
&&\quad\quad {F}{}_{\underline{M}}=(F_{\tau~{m}{n}},~F_{\tau}^m,~\bar{F}_{\tau};~
F_{\sigma}^{{m}{n}},~F_{\sigma},~F_{\sigma~m})~~,~~m=1,\cdots,5\nn\\
&&\quad\quad{\renewcommand{\arraystretch}{1.2} \left\{
\begin{array}{ccl}
	F_{\tau~{m}_1{m}_2}&=&\partial^+
 X_{m_1m_2}+\partial_{[{m}_1}{Y}_{{m}_2]}
+\frac{1}{2}\epsilon_{m_1\cdots m_5}\partial^{{m}_3}Y^{{m}_4{m}_5}
\\
F_{\tau}^m&=&-\partial^+ X^m\partial^mY-\partial_nY^{mn}\\
	\bar{F}_{\tau}&=&\partial^+ \bar{X}+\partial^{{m}}{Y}_{{m}}\\
	F_{\sigma}^{{m}_1{m}_2}&=&
	\partial^{[{m}_1}X^{{m}_2]}
	+\frac{1}{2}	\epsilon^{{m}_1\cdots {m}_5}\partial_{{m}_3}X_{{m}_4\cdots{m}_5}\\
	F_{{\sigma}~m}&=&
	\partial_m X-\partial^nX_{mn}\\
	F_{\sigma}&=&
	\partial_{{m}}X^{{m}}
\end{array}
\right.}\label{GL5Fs}
\\
\nn\\&&\rm{Gauge~transformations
}\nn\\
&&\quad\quad \kappa_{\underline{M}}=
(\kappa^{m_1m_2},~\kappa_m,~{\kappa};~\kappa_{m_1m_2},~\kappa^m,~\bar{\kappa})\nn\\
&&\quad\quad {\renewcommand{\arraystretch}{1.2}\left\{\begin{array}{ccl}
	\delta_\kappa X^m&=&
	{\partial}_n\kappa^{nm}+
	\partial^m{\kappa}\nn\\
	\delta_\kappa X_{m_1m_2}&=&
	{\partial}_{[m_1}\kappa_{m_2]}-
	\epsilon_{m_1\cdots m_5}\partial^{m_3}\kappa^{m_4m_5}
	\nn\\
	\delta_\kappa \bar{X}&=&
	\partial^m\kappa_m
\end{array}\right.}
\\
&&\quad\quad {\renewcommand{\arraystretch}{1.2}\left\{\begin{array}{ccl}
	\delta_\kappa Y_m&=&	-\partial^+{\kappa}_m
	+\partial_m\bar{\kappa}+\partial^n\kappa_{nm}
	\\
	\delta_\kappa Y^{m_1m_2}&=&\partial^+{\kappa}^{m_1m_2}
	+\partial^{[m_1}\kappa^{m_2]}+
	\frac{1}{2}\epsilon^{m_1\cdots m_5}\partial_{m_3}\kappa_{m_4m_5}
	\\
	\delta_\kappa \bar{Y}&=&\partial^+{{\kappa}}+\partial_n\kappa^{n}
\end{array}\right.}
~~~
\eea
Using with the 16$\times$16 matrix $\rho_{\underline{m}MN}$ for  $X^M=(X^m,~X_{mn},~\bar{X})$ and $Y_M=(Y_m,~Y^{mn},~Y)$ in \bref{SO55gamma} the field strengths are written as below.
\bea
&&\rm{GL(5)~field~strengths}\nn\\
&&\quad\quad {F}{}_{\underline{M}}=(F_{\tau}{}^M,~F_{\sigma~M})~~,~~M=1,\cdots,16\nn\\
&&\quad\quad{\renewcommand{\arraystretch}{1.2} \left\{\begin{array}{ccl}
{F}_\tau{}^M&=&\dot{X}^M
	-\rho^{\underline{m}MN}\partial_{\underline{m}} Y_N\\
{F}_\sigma {}_{~M}&=&\rho_{\underline{m}MN}\partial^{\underline{m}} X^N\end{array}\right.}~~~\label{GL516FS}\\
\nn\\&&\rm{Gauge~transformations
}\nn\\
&&\quad\quad \kappa_{\underline{M}}=
(\kappa_M,~
\kappa^M)\nn\\
&&\quad\quad {\renewcommand{\arraystretch}{1.2}\left\{\begin{array}{ccl}
		\delta_\kappa X^M&=&
	\rho^{\underline{m}MN}\partial_{\underline{m}} \kappa_N\\
	\delta_\kappa Y_M&=&\dot{\kappa}_M+\rho_{\underline{m}MN}\partial^{\underline{m}} \kappa^N\end{array}\right.}~~~\label{GL516kappa}
\eea

The GL(6) covariant field strength in curved background ${F}_{\underline{A}}$ is related to the one in a flat background $\don\circ{F}_{\underline{M}}$ with the vielbein field $E_{\underline{A}}{}^{\underline{M}}$
which is a coset element of SO(6,6)/SO(6;$\mathbb{C}$) as well as  the one in \bref{FSO66} as
\bea
&&G^{\underline{MN}}=
E_{\underline{A}}{}^{\underline{M}}
\don\circ{\eta}^{\underline{AB}}
E_{\underline{B}}{}^{\underline{N}}~~,~~ 
F_{\underline{A}}=E_{\underline{A}}{}^{\underline{M}}\don\circ{F}_{\underline{M}}~~~.
\eea
The number of degrees of freedom of the parameter of the coset SO(6,6)/SO(6;$\mathbb{C}$) is 
$36=25+10+1$:
25 is the number of degrees of freedom of the metric and the 3-form gauge field 
in 5 dimensions.
$10+1$ is the number of worldvolume dimensions constrained by ${\cal V}=0$, $12-1=11$.
The GL(6) covariant F10-brane action in a curved background is obtained from 
the SO(6,6) covariant F10-brane action \bref{FSO66} 
in terms of 32-component field strength $F_{\underline{M}}$ in 
\bref{FSGL6} as 
\bea
I&=&\int d^{12}\sigma~ L\nn\\
L&=&e\don\circ{F}{}_{\underline{M}}
G^{\underline{MN}} \don\circ{F}{}_{\underline{N}}
+\frac{1}{2}\lambda^{\underline{\hat{m}\hat{n}}}
\don\circ{F}{}_{\underline{M}}
(C\Sigma^{\underline{\hat{m}\hat{n}}})^{\underline{MN}} \don\circ{F}{}_{\underline{N}}
\nn\\
&=&e {F}{}_{\underline{A}}
\don\circ{\eta}{}^{\underline{AB}} {F}{}_{\underline{B}}
+\frac{1}{2}\lambda^{\underline{\hat{a}\hat{b}}}
\don\circ{F}{}_{\underline{A}}
(C\Sigma^{\underline{\hat{a}\hat{b}}})^{\underline{AB}} \don\circ{F}{}_{\underline{B}}
~~,~~\nn\\
&&\don\circ{\eta}{}^{\underline{AB}}=
\left(\begin{array}{cc}
	\hat{\eta}^{AB}
	&\\&-\hat{\eta}
	_{AB}
\end{array}\right)
\label{GL6action}
\eea
The 12-dimensional $\Sigma$ matrices  $(C\Sigma^{\underline{\hat{a}\hat{b}}})^{AB}$ are 
the same expression in \bref{12dSigma} with replacing 
$\gamma^{\underline{a}\alpha\beta}$ $\to$ $\rho^{\underline{a}AB}$, 
$\gamma^{\underline{a}}{}_{\alpha\beta}$ $\to$ $\eta^{\underline{ab}}\rho_{\underline{b}AB}$ and 
$\gamma^{\underline{ab}\alpha}{}_{\beta}$ $\to$ 
$\rho^{[\underline{a}|AC} \rho_{\underline{c}CB}\eta^{\underline{c}|\underline{b}]}$.

The GL(5) tensors  directly couple to 
 the 11-dimensional supergravity background.
The GL(5) covariant field strength in curved background ${F}{}^A$ 
is related to the one in a flat background $\don\circ{F}{}^M$
with the vielbein field $E_A{}^{M}$
which is a coset element of SO(5,5)/SO(5;$\mathbb{C}$) as well as  the one in \bref{Efs} as 
\bea
&G_{MN}=E_M{}^A
\hat{\eta}_{AB}E_N{}^B~~,~~ F{}^A=E_M{}^A\don\circ{F}{}^M
&~~~.
\eea
The Hamiltonian form action given in \bref{SO55Hamaction}
gives the same form with replacing $\gamma^{\underline{m}\mu\nu}$
with   $\rho^{\underline{m}MN}$.
The GL(5) covariant F10-brane in curved background is given  as
\bea
I&=&\int d\tau d^{10}\sigma ~L\nn\\
L&=&\varphi (g+\tilde{g})
\don\circ{F}_{+}{}^MG_{MN}
\don\circ{F}_-{}^N
-\varphi(s+\tilde{s})^{\underline{l}} 
\don\circ{F}_{+}{}^M \rho_{\underline{l}MN}
\don\circ{F}_-{}^N
\label{GL5curved}\\
&=&\varphi(g+\tilde{g}) {F}_+{}^A
\hat{\eta}_{AB}\don\circ{F}_-{}^B
-\varphi(s+\tilde{s})^{\underline{a}} 
\don\circ{F}_{+}{}^A  \rho_{\underline{a}AB}
{F}_-{}^B
\nn\\
&&\left\{\begin{array}{ccl}
	\don\circ{F}_+{}^M&=&\don\circ{F}_{\tau}{}^M
	+(\tilde{g}\hat{\eta}+\tilde{s}_{\underline{m}}\rho^{\underline{m}})^{MN}
	\don\circ{F}_{\sigma} {}_N\\
	\don\circ{F}_-{}^M&=&\don\circ{F}_{\tau}{}^M
	-({g}\hat{\eta}+{s}_{\underline{m}}\rho^{\underline{m}})^{MN}
	\don\circ{F}_{\sigma}{}_N
\end{array}\right.\nn
\eea
The GL(5) covariant F10-brane Lagrangian in curved background in terms of the selfdual and the anti-selfdual currents is given by  
\bref{SDASDFaction} as
\bea
L&=&\frac{1}{g}
\don\circ{F}_{\rm{SD}}{}^M 
G_{MN}
\don\circ{F}_{\overline{\rm{SD}}}{}^N
-\hat{\lambda}
\don\circ{F}_{\overline{\rm{SD}}}{}^M G_{MN}
\don\circ{F}_{\overline{\rm{SD}}}{}^N
-\lambda^{\underline{m}}
\don\circ{F}_{\overline{\rm{SD}}}{}^M{\rho}_{\underline{m}MN}
\don\circ{F}_{\overline{\rm{SD}}}{}^N
\label{FSDactionMN}\\
&=&\frac{1}{g}
{F}_{\rm{SD}}{}^\alpha 
\hat{\eta}_{\alpha\beta}
{F}_{\overline{\rm{SD}}}{}^\beta
-\hat{\lambda}
{F}_{\overline{\rm{SD}}}{}^\alpha \hat{\eta}_{\alpha\beta}
{F}_{\overline{\rm{SD}}}{}^\beta
-\lambda^{\underline{a}}
{F}_{\overline{\rm{SD}}}{}^\alpha{\gamma}_{\underline{a}\alpha\beta}
{F}_{\overline{\rm{SD}}}{}^\beta
\nn\\
&&\left\{\begin{array}{ccl}
	\don\circ{F}_{\rm{SD}}{}^M&=&\don\circ{F}_{\tau}{}^M
	+(g\hat{\eta}-s_{\underline{m}}\rho^{\underline{m}})^{MN}
	\don\circ{F}_{\sigma}{}_N\\
	\don\circ{F}_{\overline{\rm{SD}}}{}^M&=&\don\circ{F}_{\tau}{}^M
	-(g\hat{\eta}+s_{\underline{m}}\rho^{\underline{m}})^{MN}
	\don\circ{F}_{\sigma}{}_N
\end{array}\right.\nn
\eea

In order to couple to the 11-dimensional supergravity background 
$Y_m$ and $\bar{F}$ 
 are
rewritten in  the 5-dimensional dual as
\bea
&Y_m=\frac{1}{4!}\epsilon_{mn_1\cdots n_4}\bar{Y}^{n_1\cdots n_4}
~~,~~
\bar{F}
=\frac{1}{5!}F
{}_{m_1\cdots m_5}\epsilon^{m_1\cdots m_5}
\eea
The background of SO(5,5) vielbein 
given in \bref{EEC3} is given by
the metric $e_m{}^a$ and the three form gauge field $C_{m_1m_2m_3}^{[3]}$ in the 5-dimensional space $m=1,\cdots,5$. 
The field strength $\don\circ{F}{}^M$ 
including both   $\don\circ{F}_{\tau}{}^M$ and $\don\circ{F}_{\sigma}{}^M$ 
in the backgrounds are as follows. 
\bea
&&F{}^{A}~=~E_{M}{}^{A}\don\circ{F}
{}^{M}~~~\nn\\.
&&
{\renewcommand{\arraystretch}{1.6}\left\{\begin{array}{ccl}
F{}^a&=&F
{}^me_m{}^a\\
F{}_{a_1a_2}&=&-F
{}^mC^{[3]}{}_{mn_1n_2}e_{a_1}{}^{n_1}e_{a_2}{}^{n_2}
+F{}_{m_1m_2}e_{a_1}{}^{m_1}e_{a_2}{}^{m_2}\\
F{}_{a_1\cdots a_5}&=&
+\frac{1}{2\cdot 3!}F
{}_{m_1m_2}C^{[3]}{}_{m_3m_4m_5}e_{[a_1}{}^{m_1}\cdots e_{a_5]}{}^{m_5}\\
&&+F{}_{m_1\cdots m_5}e_{a_1}{}^{m_1}\cdots e_{a_5}{}^{m_5}\end{array}\right.}
\label{SUGRAbg}
\eea

\par
\vskip 6mm
\section{Perturbative M-theory 5-brane action}
In order to obtain an action for the perturbative M-theory 5-brane 
coupled to the 11-dimensional supergravity background,
we preserve the number of the SO(5,5) currents. 
The worldvolume dimensions of F-theory  is reduced by solving ${\cal V}=
\partial_{\underline{m}}\eta^{\underline{m}\underline{n}}\partial_{\underline{n}}=0$ as $\bar{\partial}_m=\frac{1}{4!}\epsilon_{m_1\cdots m_5}\partial^{m_2\cdots m_5}=0$ consistently. 
The 5-dimensional worldvolume theory is obtained by the following sectioning \cite{Siegel:2020qef}
\bea
\bar{\partial}_m=0~\to~{\cal V}=2\partial^m\bar{\partial}_m=0~~.\label{5wv}
\eea

The selfdual and the anti-selfdual currents for the M5-brane are given as
\bea
&&{\renewcommand{\arraystretch}{1.6}\left\{\begin{array}{ccl}
\dd_m&=& P_m+\partial^n X_{mn}\\
\dd^{m_1m_2}&=& P^{m_1m_2}+\partial^{[m_2} X^{m_1]}\label{SDCdualM5}\\
\bar{\dd}&=& \bar{P}\end{array}\right.}
 \\
&&{\renewcommand{\arraystretch}{1.6}\left\{\begin{array}{ccl}
\tilde{\dd}_m&=& P_m-\partial^n X_{mn}\nn\\
\tilde{\dd}^{m_1m_2}&=& P^{m_1m_2}-\partial^{[m_2} X^{m_1]}\label{ASDCdualM5}\\
\bar{\tilde{\dd}}&=& \bar{P}\end{array}\right.} ~~~.
\eea
The SO(5,5) current algebras in \bref{CA3} is reduced to the subalgebra
which is the same one for the SL(5) case \cite{Hatsuda:2012vm}
\bea
&&{\renewcommand{\arraystretch}{1.6}\left\{\begin{array}{ccl}
	\left[\dd_m(\sigma),\dd^{n_1n_2}(\sigma')\right]&=&2i\delta_m^{[n_1}\partial^{n_2]}\delta(\sigma-\sigma')\\
\left[\dd_m(\sigma),\bar{\dd}(\sigma')\right]&=&0\\
\left[\dd^{m_1m_2}(\sigma),\dd^{n_3n_4}(\sigma')\right]&=&0
\end{array}\right.}\\
&&\left[\tilde{\dd}_\mu(\sigma),\dd_\nu(\sigma')\right]=0\nn\\
&&{\renewcommand{\arraystretch}{1.6}\left\{\begin{array}{ccl}
	\left[\tilde{\dd}_m(\sigma),\tilde{\dd}^{n_1n_2}(\sigma')\right]&=&-2i\delta_m^{[n_1}\partial^{n_2]}\delta(\sigma-\sigma')\\
\left[\tilde{\dd}_m(\sigma),\bar{\tilde{\dd}}(\sigma')\right]&=&0\\
\left[\tilde{\dd}^{m_1m_2}(\sigma),\tilde{\dd}^{n_3n_4}(\sigma')\right]&=&0\end{array}\right.}\nn
\eea
A set of the Virasoro constraints and the Gauss law constraints in
\bref{Virasoro2} and \bref{Ucon}
 are reduced to
\bea
&&{\renewcommand{\arraystretch}{1.6}\left\{\begin{array}{ccl}
{\cal S}^{m}&=&\frac{1}{2}\dd_n\dd^{nm}=0\\
\bar{\cal S}_{m}&=&\frac{1}{2}[\dd_n\bar{\dd}+\frac{1}{8}\epsilon_{mm_1\cdots m_4}\dd^{m_1m_2}\dd^{m_3m_4}]=0\\
{\cal T}&=&\frac{1}{2}[\dd_m\eta^{mn}\dd_n
+\frac{1}{2}\dd^{m_1m_2}\eta_{m_1n_2}\eta_{m_2n_2}\dd^{n_1n_2}
+\bar{\dd}^2]=0
\end{array}\right.}\nn\\
&&{\renewcommand{\arraystretch}{1.6}\left\{\begin{array}{ccl}
	{\cal U}&=&\dd_m \partial^m=0\\
{\cal U}^m&=&\bar{\dd}\partial^m=0\\
{\cal U}_{m_1m_2}&=&
\epsilon_{m_1\cdots m_5} \dd^{m_3m_4}\partial^{m_5}
=0\\
{\cal V}&=&0~~~.
\end{array}\right.}
\eea

An action for the M-theory 5-brane  in curved background is given from the  F-theory 10-brane \bref{Fcurvepmaction} by sectioning the worldvolume into 5 dimensions as
\bea
I&=&\displaystyle \int d\tau d^5\sigma ~L\nn\\
L&=&
\varphi\don\circ{F}_{+}{}^M\left\{
(g+\tilde{g})
 G_{MN}
-\varphi(s+\tilde{s})^{\underline{l}} 
\rho_{\underline{l}MN}\right\}
\don\circ{F}_-{}^N
\label{GL5curved1}\\
&=&\varphi {F}_+{}^A\left\{
(g+\tilde{g})
\hat{\eta}_{AB}
-\varphi(s+\tilde{s})^{\underline{a}} 
\rho_{\underline{a}AB}\right\}
{F}_-{}^B~~~.
\nn
\eea

 Now let us construct the action for the M-theory 5-brane 
in terms of the selfdual and the anti-selfdual currents
where the anti-selfdual currents are auxiliary introduced to make a free kinetic term.
The selfdual (SD) and the anti-selfdual ($\overline{\rm{SD}}$) field strengths are given by
\bea
&&\don\circ{F}_{\rm{SD}/\overline{\rm{SD}}}{}^M=(\don\circ{F}_{\rm{SD}/\overline{\rm{SD}}}{}^m,~
\don\circ{F}_{\rm{SD}/\overline{\rm{SD}}}{}_{~{m}{n}},~\don\circ{F}_{\rm{SD}/\overline{\rm{SD}}})\nn\\
&&\left\{
\begin{array}{ccl}
\don\circ{F}_{\rm{SD}/\overline{\rm{SD}}}{}^m&=&\don\circ{F}_{\tau}{}^m\pm
\left\{g\eta^{mn}\don\circ{F}_{\sigma~}{}_n
-(s_n\don\circ{F}_{\sigma}{}^{mn}+\bar{s}^m \don\circ{F}_{\sigma})
\right\}
\\\don\circ{F}_{\rm{SD}/\overline{\rm{SD}}}{}_{~{m_1}{m_2}}&=&
\don\circ{F}_{\tau}{}_{~m_1m_2}\pm
\left\{g\eta_{m_1n_1}\eta_{m_2n_2}\don\circ{F}{}_{\sigma}{}^{n_1n_2}
+(s_{[m_1}\don\circ{F}_{\sigma|m_2]}-
\epsilon_{m_1m_2n_1n_2k}\bar{s}^k \don\circ{F}_{\sigma}{}^{n_1n_2})
\right\}
\\\don\circ{F}_{\rm{SD}/\overline{\rm{SD}}}&=&
\don\circ{\bar{F}}_{\tau}\pm
\left\{g\don\circ{F}_{\sigma}
-\bar{s}^n\don\circ{F}_{\sigma~n}\right\}
\end{array}
\right.\nn\\
\eea
where the selfdual or the anti-selfdual currents picks up the $+$ or $-$ among $\pm$ sign respectively.
The GL(5) covariant field strengths in flat space are given by
\bea
&&\don\circ{F}{}_\tau{}{^M}=(\don\circ{F}_{\tau}{}^m,~\don\circ{F}_{\tau~{m}{n}},~\don\circ{\bar{F}}_{\tau})
\nn\\
&&{\renewcommand{\arraystretch}{1.2}\left\{
\begin{array}{ccl}
	\don\circ{F}_\tau^m&=&-
	\dot{X}^m+\partial^mY
\\
	\don\circ{F}_{\tau~{m}_1{m}_2}&=&
	\dot{X}_{m_1m_2}
	+\frac{1}{2}\epsilon_{m_1\cdots m_5}\partial^{{m}_3}Y^{{m}_4{m}_5}
	\\
\don\circ{	\bar{F}}_\tau&=&
\dot{\bar{X}}+\partial^{{m}}{Y}_{{m}}
\end{array}
\right.}\\
&&\don\circ{F}{}_{\sigma~M}=(\don\circ{F}_{\sigma;m},~\don\circ{F}_\sigma{}^{{m}{n}},~\don\circ{F}_\sigma)
\nn\\
&&{\renewcommand{\arraystretch}{1.2}\left\{
\begin{array}{ccl}
	\don\circ{F}_{\sigma;m}&=&\partial^nX_{nm}\\
	\don\circ{F}_\sigma^{{m}_1{m}_2}&=&\partial^{[{m}_1}X^{{m}_2]}
	\\
	\don\circ{F}_\sigma&=&0
\end{array}
\right.}
\eea
where  $\partial^+$ is replaced with the $\tau$ derivative denoted by $\dot{X}^M$.
The SO(5,5) background $G_{MN}=E_M{}^A\hat{\eta}_{AB}E_N{}^B$ is given by the vielbein
$E_M{}^A$  from \bref{EEC3} as
\bea
E_M{}^A&=&\left(
\begin{array}{ccc}
	e_m{}^a&-C^{[3]}_{mn_1n_2}e_{a_1}{}^{n_1}e_{a_2}{}^{n_2}&
	\frac{1}{2\cdot 3!}
	C^{[3]}_{m[n_1n_2}C^{[3]}_{n_3n_4n_5]}e_{a_1}{}^{n_1}\cdots e_{a_5}{}^{n_5}
	\\
	0&e_{[a_1}{}^{m_1}e_{a_2]}{}^{m_2}&
	\frac{1}{3!}C^{[3]}_{m_3m_4m_5}e_{[a_1}{}^{m_1}
	\cdots e_{a_5]}{}^{m_5}\\
	0&0&e_{[a_1}{}^{m_1}\cdots e_{a_5]}{}^{m_5}
\end{array}\right)\label{spGG}~~~.
\eea
We propose a perturbative action for a M-theory 5-brane in the curved background in terms of the selfdual and the anti-selfdual currents in \bref{FSDactionMN}
as follow.
\bea
I&=&\displaystyle \int d\tau d^5\sigma ~L\nn\\
L&=&\frac{1}{g}
\don\circ{F}_{\rm{SD}}{}^M 
G_{MN}
\don\circ{F}_{\overline{\rm{SD}}}{}^N
-\hat{\lambda}
\don\circ{F}_{\overline{\rm{SD}}}{}^M G_{MN}
\don\circ{F}_{\overline{\rm{SD}}}{}^N
-\lambda^{\underline{m}}
\don\circ{F}_{\overline{\rm{SD}}}{}^M{\rho}_{\underline{m}MN}
\don\circ{F}_{\overline{\rm{SD}}}{}^N
\label{FSDactionMNbf}\\
&=&\frac{1}{g}
{F}_{\rm{SD}}{}^A
\hat{\eta}_{AB}
{F}_{\overline{\rm{SD}}}{}^B
-\hat{\lambda}
{F}_{\overline{\rm{SD}}}{}^A \hat{\eta}_{AB}
{F}_{\overline{\rm{SD}}}{}^B
-\lambda^{\underline{a}}
{F}_{\overline{\rm{SD}}}{}^A{\rho}_{\underline{a}AB}
{F}_{\overline{\rm{SD}}}{}^B
\nn
\eea
The first term is a free kinetic term for GL(5) tensor fields 
on a (5+1)-dimensional worldvolume, while the rest is 
constraints of the anti-selfdual currents.

\section{Conclusions}

We have presented the F-theory 10-brane actions with
the SO(5,5) U-duality symmety and  the SO(6,6) enlarged U-duality symmetry.
At first the SO(5,5) current algebras on the 10-dimensional worldvolume 
are presented in both the SO(5,5) spinor representation and the GL(5) tensor representation.
The former reveals the gauge symmetry structure generated by the Gauss law constraint
where the 10-dimensional worldvolume translation and the 16-dimensional 
spacetime current satisfy the bosonic $\kappa$-symmetry structure. 
The latter gives direct coupling to the 11-dimensional supergravity background fields.
Next the action of the F-theory 10-brane is obtained by the Legendre transformation of the Hamiltonian constructed by the set of
Virasoro constraints.
Applying the double zweibein method to F-theory allows to give 
the 10-dimensional worldvolume covariant actions.
Then we have also constructed the F-theory 10-brane action with the SO(6,6) symmetry 
in the Lagrangian formalism.
The worldvolume is enlarged to 12 dimensional brane spacetime,
while the target spacetime is enlarged to the 32 dimensional spacetime.
The background vielbein represents the coset SO(6,6)/SO(6;$\mathbb{C}$)
 including both the spacetime background SO(5,5)/SO(5;$\mathbb{C}$)
and the worldvolume vielbein.

We have also presented the action for the perturbative M-theory 5-brane in curved spacetime 
by sectioning the worldvolume of the F-theory 10-brane action. 
The spacetime is 16 dimensional
manifesting the SO(5,5) background coupling.
The action is sum of the free kinetic term
and the bilinears of the selfduality constraint. 
So now quantization of the F10-brane and the M5-brane is challenging problem.
It is also interesting to note that 5-brane is the only object that appears in common in all theories;
  type I, IIA, IIB superstrings, SO(32), E$_8\times$E$_8$ heterotic superstrings, M-theory, F-theory.
In F-theory the 5-brane represents the SL(5) U-duality symmetry \cite{Linch:2015fya}.
Recently it was shown that current algebras of 5-branes are preserved
under the S and T-duality transformations with renaming the spacetime coordinates, 
where 5-branes include the NS5-brane, the D5-brane,
KK5-branes and exotic 5-branes in 32-supersymmetric string theories \cite{Hatsuda:2020buq}.
5-brane may give a clue of duality web including 16 supersymmetric string theories.

Many interesting topics are unsolved such as
supersymmetric actions of F-theory and M-theory,
first quantization of branes and spectrum, amplitudes,
and duality web including 16-supersymmetric theories.

\section*{Acknowledgements}

We would like to thank Di Wang for useful discussions.
M.H. would like to thank the Simons Center for Geometry and Physics for
hospitality during ``the 2019 Summer Simons workshop in Mathematics and Physics"
where this work has been developed.
W.S. is supported by NSF grant PHY-1915093.

\appendix
\section{Indices}
\label{ind}

Indices are summarized as follows.

\begin{center}
	\begin{tabular}{cccc}
	&number~of~d.o.f.&~~~curved~~~&~~~~flat~~~~\\\hline
	F-theory&&&
	\\\hline
spacetime&&&\\
SO(5,5)~spinor&16&$\mu,\nu,\cdots$&
$\alpha,\beta,\cdots$\\
SO(6,6)~spinor&32&$\underline{\mu},\underline{\nu},\cdots$&
$\underline{\alpha},\underline{\beta},\cdots$\\
GL(5)~tensor&16&$M,N,\cdots$~~&$A,B,\cdots$\\
GL(6)~tensor&32&$\underline{M},\underline{N},\cdots$&$\underline{A},\underline{B},\cdots$~\\\hline
worldvolume&&&\\
SO(5,5)~vector&10&$\underline{m},\underline{n},\cdots$&
$\underline{a},\underline{b},\cdots$\\
SO(6,6)~vector&12&$\underline{\hat{m}},\underline{\hat{n}},\cdots$&
$\underline{\hat{a}},\underline{\hat{b}},\cdots$\\
GL(5)~vector&5&$m,n,\cdots$&$a,b,\cdots$\\
GL(6)~vector&6&$\hat{m},\hat{n},\cdots$&$\hat{a},\hat{b},\cdots$\\\hline\hline
T-theory&&&\\
(only $\S$4-1)&&&\\\hline
spacetime&&&\\
O($D,D$)~vector&$2D$&$M,N,\cdots$&
$A,B,\cdots$\\
left-handed&$D$&$\overline{M},\overline{N},\cdots$&
$\overline{A},\overline{B},\cdots$\\
right-handed&$D$&$\underline{M},\underline{N},\cdots$&
$\underline{A},\underline{B},\cdots$\\\hline
worldvolume&&&\\
SO(1,1)~vector&2&${m},{n},\cdots$&
${a},{b},\cdots$\\\hline
\end{tabular}
\end{center}
\par

\vskip 6mm

\section{Brackets}
\label{brackets}
In the F-theory spacetime the Lie derivative is modified
in such a way that it is the SO(5,5) U-duality symmetry covariant.
We compute  commutators in the SO(5,5) spinor representation 
which is easier than the GL(5) representation.
For vector functions  $V_i{}^{\mu}(X)$ with $i=1,2$ in the 16-dimensional spacetime
a commutator brackets of these vectors  is given by
\bea
\left[V_1^\mu\dd_\mu(\sigma),V_2^\nu\dd_\nu(\sigma')\right]&=&
2i\left(\frac{1-K}{2}\Phi_{(12)}^{\underline{m}}(\sigma)+\frac{1+K}{2}
\Phi_{(12)}^{\underline{m}}(\sigma')\right)\partial_{\underline{m}}\delta(\sigma-\sigma')\nn\\
&&-i\left(V_1^\nu\partial_\nu V_2^\mu - V_2^\nu\partial_\nu V_1^\mu \right)\dd_\mu\delta(\sigma-\sigma')\\
&&+i\left(\frac{1-K}{2}V_1\gamma^{\underline{m}}\partial_\mu V_2 -\frac{1+K}{2}\partial_\mu V_1 \gamma^{\underline{m}}V_2\right)(\gamma_{\underline{m}}\dd)^\mu\delta(\sigma-\sigma')~~\nn\\
\Phi_{(12)}^{\underline{m}}&=&V_1{}^\mu\gamma^{\underline{m}}{}_{\mu\nu}V_2{}^{\nu}\nn~~~.
\eea 
The exceptional Courant bracket is given by  $K=0$ as
\bea
\left[V_1^\mu\dd_\mu(\sigma),V_2^\nu\dd_\nu(\sigma')\right]&=&
{i}\left(\Phi_{(12)}^{\underline{m}}(\sigma)+\Phi_{(12)}^{\underline{m}}(\sigma')
\right)\partial_{\underline{m}}\delta(\sigma-\sigma')
-iV_{[12]}{}^\mu\dd_\mu\delta(\sigma-\sigma')\nn\\
V_{[12]}{}^\mu&=&V_{[1|}^\nu\partial_\nu V_{|2]}^\mu 
-\frac{1}{2}(V_{[1|}{}^\rho\gamma^{\underline{m}}{}_{\rho\lambda}\partial_\nu V_{|2]}{}^\lambda )\gamma_{\underline{m}}{}^{\nu\mu}~~\label{ExCou},
\eea
while the exceptional Dorfman bracket is given by $K=1$ as
\bea
\left[V_1^\mu\dd_\mu(\sigma),V_2^\nu\dd_\nu(\sigma')\right]&=&
2i\Phi_{(12)}^{\underline{m}}(\sigma')
\partial_{\underline{m}}\delta(\sigma-\sigma')\label{ExDor}\\
&&-i\left(
V_{[1|}^\nu\partial_\nu V_{|2]}^\mu +
(\partial_\nu V_1 \gamma^{\underline{m}}V_2) \gamma_{\underline{m}}{}^{\nu\mu}
\right)\dd_\mu\delta(\sigma-\sigma')
~~.\nn
\eea 

\par
\vskip 6mm
\section{11-dimensional tensor representation}
\label{11dtensor}

The 16 component SO(5,5) spinor current
is decomposed under the GL(5) as $16 \to 5 \oplus 10 \oplus 1$.
We present the current algebras preserving the full tensor indices such as  $\dd^{m_1\cdots m_5}$ with $m=1,\cdots,5$ in order to manifest the 11-dimensional supergravity background.
The SO(5,5) current 
for the M5-brane  is obtained as
$\dd_M=(\dd_m, \dd^{m_1m_2},\dd^{m_1\cdots m_5})$ \cite{}. 
The F-theory SO(5,5) current algebras give in \bref{currentflat} 
are rewritten in terms of the GL(5) tensors as
\bea
&&\left\{\begin{array}{rcl}
	\left[\dd_m(\sigma),\dd^{n_1n_2}(\sigma')\right]&=&2i\delta_m^{[n_1}\partial^{n_2]}\delta(\sigma-\sigma')\nn\\
	\left[\dd_m(\sigma),\dd^{n_1\cdots n_5}(\sigma')\right]&=&\frac{2i}{4!}\delta_m^{[n_1}\partial^{n_2 \cdots n_5]}\delta(\sigma-\sigma')\label{CA}\\
	\left[\dd^{m_1m_2}(\sigma),\dd^{n_3n_4}(\sigma')\right]&=&2i\partial^{m_1m_2n_3n_4}\delta(\sigma-\sigma')\nn\end{array}\right.\\
&&~~~~~~~~~~\left[{\dd}_M(\sigma),\tilde{\dd}_N(\sigma')\right]~=~0\nn 
\\
&&\left\{\begin{array}{rcl}\left[\tilde{\dd}_m(\sigma),\tilde{\dd}^{n_1n_2}(\sigma')\right]&=&-2i\delta_m^{[n_1}\partial^{n_2]}\delta(\sigma-\sigma')\nn\\
	\left[\tilde{\dd}_m(\sigma),\tilde{\dd}^{n_1\cdots n_5}(\sigma')\right]&=&-\frac{2i}{4!}\delta_m^{[n_1}\partial^{n_2 \cdots n_5]}\delta(\sigma-\sigma')\label{CA2}\\
	\left[\tilde{\dd}^{m_1m_2}(\sigma),\tilde{\dd}^{n_3n_4}(\sigma')\right]&=&-2i\partial^{m_1m_2n_3n_4}\delta(\sigma-\sigma')\nn\end{array}\right.
\eea
where the $\delta(\sigma)$ stands for the 10-dimensional worldvolume function $\delta^{(10)}(\sigma-\sigma')$.
The bosonic  coordinates, 
$X^M=(X^m,X_{m_1m_2},X_{m_1\cdots m_5} )$ and $P_M=(P_m,P^{m_1m_2},P^{m_1\cdots m_5})$, satisfy the following canonical commutators 
\bea
\left\{\begin{array}{rcl}
	\left[P_m(\sigma),X^n(\sigma)\right]&=&\frac{1}{i}\delta_m^n\delta(\sigma-\sigma')\\
	\left[P^{m_1m_2}(\sigma),X_{n_1n_2}(\sigma)\right]&=&\frac{1}{i}\delta_{m_1}^{[n_1}
	\delta_{m_2}^{n_2]}\delta(\sigma-\sigma')\label{cancom}\\
	\left[P^{m_1\cdots m_5}(\sigma),X_{n_1\cdots n_5}(\sigma)\right]&=&\frac{1}{i}\delta_{m_1}^{[n_1}\cdots 
	\delta_{m_5}^{n_5]}\delta(\sigma-\sigma')\end{array}\right. ~~~.
\eea

The covariant derivatives, which are selfdual currents, are given as
\bea
\left\{\begin{array}{rcl}
	\dd_m&=& P_m+\partial^n X_{mn}+\frac{1}{4!}\partial^{m_1 \cdots m_4}X_{mm_1\cdots m_4}\\
	\dd^{m_1m_2}&=& P^{m_1m_2}-\partial^{[m_1} X^{m_2]}+
	\frac{1}{2}\partial^{m_1 \cdots m_4}X_{m_3 m_4}\label{SDC}\\
	\dd^{m_1\cdots m_5}&=& P^{m_1\cdots m_5}
	+\frac{1}{4!}\partial^{[m_1 \cdots m_4}X^{m_5]}\end{array}\right. ~~~,
\eea
and the symmetry generators, which are anti-selfdual currents, are given as 
\bea
\left\{\begin{array}{rcl} 
	\tilde{\dd}_m&=& P_m-\partial^n X_{mn}-\frac{1}{4!}\partial^{m_1 \cdots m_4}X_{mm_1\cdots m_4}\\
	\tilde{\dd}^{m_1m_2}&=& P^{m_1m_2}+\partial^{[m_1} X^{m_2]}-
	\frac{1}{2}\partial^{m_1 \cdots m_4}X_{m_3 m_4\label{ASDC}}\\
	\tilde{\dd}^{m_1\cdots m_5}&=& P^{m_1\cdots m_5}
	-\frac{1}{4!}\partial^{[m_1 \cdots m_4}X^{m_5]}\end{array}\right. ~~~.
\eea
The  Virasoro constraints are the followings 
\bea
\left\{\begin{array}{rcl}
	{\cal S}^m&=&\frac{1}{2}\dd_n \dd^{nm}=0\\
	{\cal S}^{m_1\cdots m_4}&=&\frac{1}{2}\left[\dd_n \dd^{nm_1\cdots m_4}
	+\frac{1}{8}\dd^{[m_1m_2}\dd^{m_3m_4]}\right]
	=0\label{Virasoro1}\\
	{\cal T}&=&\frac{1}{4}\left[\dd_m\delta^{mn}\dd_n+\frac{1}{2} \dd^{m_1m_2}\delta_{m_1n_1}\delta_{m_2n_2}\dd^{n_1n_2}\right.\nn\\
	&&\left.+\frac{1}{5!}\dd^{m_1\cdots m_5}\delta_{m_1n_2}\cdots\delta_{m_5n_5}
	\dd^{n_1\cdots n_5}\right]
	=0\end{array}\right.
\eea

For the tensor calculation there are several useful relations to obtain the Virasoro algebras
for the totally antisymmetric tensors $t^{m_1m_2}$ and $T^{m_1\cdots m_4}$
which are obtained by the totally anti-symmetric 6 indices as
\bea
\frac{1}{3!}V^{[m_1}\epsilon^{m_2m_3m_4]n_1n_2}&=&-V^{[n_1}\epsilon^{n_2]m_1m_2m_3m_4}\nn\\
\frac{1}{4}t^{[m_1m_2}T^{m_3m_4]n_1n_2}&=&-t^{n_1n_2}T^{m_1\cdots m_4}\nn\\
\frac{1}{4}\epsilon^{ml_1l_2[n_1n_2}T^{n_3n_4]}&=&\epsilon^{mn_1\cdots n_4}T^{l_1l_2}~~~.
\label{useful}
\eea
\par
\vskip 6mm
\section{Double vielbein formulation}
\label{2vb}

The double vielbein formulation in the simplest example is explain in this appendix.
The selfduality constraint in the T-theory Hamiltonian with a flat worldsheet
is the anti-selfduality current is 0:
The 2D-dimensional selfdual current (the covariant derivative) 
and   the 2D-dimensional selfduality current (the symmetry generator current) 
 are given by
\bea
\left\{\begin{array}{ccl}
\dd_M&=&P_M+\partial_\sigma X^N\eta_{NM}\\
\tilde{\dd}_M&=&P_M-\partial_\sigma X^N\eta_{NM}
\end{array}\right.
~~~.
\eea
When the 2D-dimensional coordinate $X^M$ is written in terms of the D-dimensional coordinates as $(x,y)$ and the canonical conjugates as $(p_x,p_y)$
for the O($D,D$) invariant metric 
$\eta_{MN}=\left(\begin{array}{cc}0&1\\1&0\end{array}\right)$, 
the currents are written as 
\bea
&\dd_M=\left\{\begin{array}{ccl}p_x+\partial_\sigma y\\
p_y+\partial_\sigma x\end{array}\right.
~~,~~
\tilde{\dd}_M=\left\{\begin{array}{ccl}p_x-\partial_\sigma y\\
	p_y-\partial_\sigma x\end{array}\right.&~~~.
\eea 
The selfduality constraint is the anti-selfdual current is 0 $\tilde{\dd}_M=0$
in the usual formulation.
By using the selfduality constraint  $p_y=\partial_\sigma x$,
the selfdual current reduces into the D-dimensional momenta and the winding modes
$\dd_M\to (p_x, \partial_\sigma x)$.

When the Hamiltonian is made from only the selfdual currents,
 the Hamiltonian form Lagrangian gives a chiral scalar Lagrangian where the term $(\partial_\sigma x){}^2$ is absent.
\bea
I&=&\int L~~,~~L=\dot{x}p-H~~\nn\\
H&=&\frac{g}{2}\dd{}^2
~~
\to L=\frac{1}{g}(\dot{X}^2+2g\dot{X}~\partial_\sigma X)~~~.\nn
\eea
 But adding the squared anti-selfdual current as a constraint 
  with the Lagrange multiplier $\tilde{g}$ leads to the worldsheet covariant action as
 \bea
H&=&\frac{g}{4}\dd{}^2
+\frac{\tilde{g}}{4}\tilde{\dd}{}^2
\to~L=\frac{1}{g+\tilde{g}}(\dot{X}-g\partial_\sigma X)(\dot{X}+\tilde{g} \partial_\sigma X)~~~.
\eea 
This is rewritten in terms of the selfdual and the anti-selfdual currents as
\bea
&L=\displaystyle\frac{1}{2g}J_{\rm{SD}}J_{\overline{\rm{SD}}}+
\frac{g-\tilde{g}}{2g(g+\tilde{g})}J_{\overline{\rm{SD}}}{}^2&\label{Jtauonly}\\
&
\left\{\begin{array}{ccl}
J_{\rm{SD}}{}^M&=&\dot{X}^M+g\hat{\eta}^{MN}\eta_{NL}\partial_\sigma X^L\\
J_{\overline{\rm{SD}}}&=&\dot{X}^M-g\hat{\eta}^{MN}\eta_{NL}\partial_\sigma X^L
\end{array}\right.&\nn\\
&J_{\rm{SD}}{}^M=
\left\{\begin{array}{l}
\dot{x}+g\partial_\sigma y\\\dot{y}+{g} \partial_\sigma x
\end{array}\right.~~,~~
J_{\overline{\rm{SD}}}=
\left\{\begin{array}{l}
	\dot{x}-g\partial_\sigma y\\\dot{y}-{g} \partial_\sigma x
\end{array}\right.
&~~~.\label{ASDcon}
\eea 
The first term in the Lagrangian is the free kinetic term 
while the second term is the selfduality constraint in 
a bilinear form.
The bilinear form constraint reduces into the anti-selfdual current to be 0,
which relates the doubled coordinates $x$ and $y$ as the usual selfduality constraint.



\small
\linespread{1.1}\selectfont
\raggedright

\providecommand{\href}[2]{#2}\begingroup\raggedright\endgroup


\begin{thebibliography}{10}



\bibitem{Siegel:1993bj}
W.~Siegel, ``{Manifest duality in low-energy superstrings},'' in {\em
	{International Conference on Strings 93 Berkeley, California, May 24-29,
		1993}}, pp.~353--363.
\newblock 1993.
\newblock
\href{http://arxiv.org/abs/hep-th/9308133}{{\ttfamily arXiv:hep-th/9308133
		[hep-th]}}.
\newblock

\bibitem{Siegel:1993th}
W.~Siegel, ``{Superspace duality in low-energy superstrings},''
\href{http://dx.doi.org/10.1103/PhysRevD.48.2826}{{\em Phys. Rev.} {\bfseries
		D48} (1993) 2826--2837},
\href{http://arxiv.org/abs/hep-th/9305073}{{\ttfamily arXiv:hep-th/9305073
		[hep-th]}}.

\bibitem{Siegel:1993xq}
W.~Siegel, ``{Two vierbein formalism for string inspired axionic gravity},''
\href{http://dx.doi.org/10.1103/PhysRevD.47.5453}{{\em Phys. Rev.}
	 {\bfseries	D47} (1993) 5453--5459},
\href{http://arxiv.org/abs/hep-th/9302036}
{{\ttfamily arXiv:hep-th/9302036		[hep-th]}}.


\bibitem{Polacek:2013nla}
M.~Pol\'{a}{\v{c}}ek and W.~Siegel, ``{Natural curvature for manifest T-duality},''
\href{http://dx.doi.org/10.1007/JHEP01(2014)026}{{\em JHEP} {\bfseries 01}
	(2014) 026},
\href{http://arxiv.org/abs/1308.6350}{{\ttfamily arXiv:1308.6350 [hep-th]}}.

\bibitem{Hatsuda:2014aza}
M.~Hatsuda, K.~Kamimura, and W.~Siegel, ``{Ramond-Ramond gauge fields in
	superspace with manifest T-duality},''
\href{http://dx.doi.org/10.1007/JHEP02(2015)134}{{\em JHEP} {\bfseries 02}
	(2015) 134},
\href{http://arxiv.org/abs/1411.2206}{{\ttfamily arXiv:1411.2206 [hep-th]}}.

\bibitem{Hatsuda:2014qqa}
M.~Hatsuda, K.~Kamimura, and W.~Siegel, ``{Superspace with manifest T-duality
	from type II superstring},''
\href{http://dx.doi.org/10.1007/JHEP06(2014)039}{{\em JHEP} {\bfseries 06}
	(2014) 039},
\href{http://arxiv.org/abs/1403.3887}{{\ttfamily arXiv:1403.3887 [hep-th]}}.

\bibitem{Polacek:2014cva}
M.~Pol\'{a}{\v{c}}ek and W.~Siegel, ``{T-duality off shell in 3D Type II superspace},''
\href{http://dx.doi.org/10.1007/JHEP06(2014)107}{{\em JHEP} {\bfseries 06}
	(2014) 107},
\href{http://arxiv.org/abs/1403.6904}{{\ttfamily arXiv:1403.6904 [hep-th]}}.

\bibitem{Hatsuda:2015cia}
M.~Hatsuda, K.~Kamimura, and W.~Siegel, ``{Type II chiral affine Lie algebras
	and string actions in doubled space},''
\href{http://dx.doi.org/10.1007/JHEP09(2015)113}{{\em JHEP} {\bfseries 09}
	(2015) 113},
\href{http://arxiv.org/abs/1507.03061}{{\ttfamily arXiv:1507.03061 [hep-th]}}.

\bibitem{Hatsuda:2017tfa}
M.~Hatsuda, K.~Kamimura, and W.~Siegel, ``{Manifestly T-dual formulation of AdS
	space},''
\href{http://dx.doi.org/10.1007/JHEP05(2017)069}{{\em JHEP}
	{\bfseries 05} (2017) 069},
\href{http://arxiv.org/abs/1701.06710}{{\ttfamily arXiv:1701.06710 [hep-th]}}.

\bibitem{Hatsuda:2018tcx}
M.~Hatsuda and W.~Siegel,
``O(D, D) gauge fields in the T-dual string Lagrangian,''
\href{http://dx.doi:10.1007/JHEP02(2019)010}{{\em JHEP}
	{\bfseries 02} (2019) 010},
\href{http://arxiv.org/abs/1810.04761}{{\ttfamily arXiv:1810.04761 [hep-th]}}.

\bibitem{Hatsuda:2019xiz}
M.~Hatsuda and W.~Siegel,
``T-dual Superstring Lagrangian with double zweibeins,''
\href{http://dx.doi:10.1007/JHEP03(2020)058}{{\em JHEP}
	{\bfseries 03} (2020) 058},
\href{http://arxiv.org/abs/1912.05092}{{\ttfamily arXiv:1912.05092 [hep-th]}}.






\bibitem{Hull:2009mi}
C.~Hull and B.~Zwiebach,
``Double Field Theory,''
\href{http://dx.doi:10.1088/1126-6708/2009/09/099}{{\em JHEP} {\bfseries
		09} (2009) 099},
\href{http://arxiv.org/abs/0904.4664}{{\ttfamily arXiv:0904.4664 [hep-th]}}.


\bibitem{Hull:2009zb}
C.~Hull and B.~Zwiebach, ``{The Gauge algebra of double field theory and
	Courant brackets},''
\href{http://dx.doi.org/10.1088/1126-6708/2009/09/090}{{\em JHEP} {\bfseries
		09} (2009) 090},
\href{http://arxiv.org/abs/0908.1792}{{\ttfamily arXiv:0908.1792 [hep-th]}}.

\bibitem{Zwiebach:2011rg}
B.~Zwiebach, ``{Double Field Theory, T-Duality, and Courant Brackets},''
\href{http://dx.doi.org/10.1007/978-3-642-25947-0_7}{{\em Lect. Notes Phys.}
	{\bfseries 851} (2012) 265--291},
\href{http://arxiv.org/abs/1109.1782}{{\ttfamily arXiv:1109.1782 [hep-th]}}.

\bibitem{Berman:2013eva}
D.~S. Berman and D.~C. Thompson, ``{Duality Symmetric String and M-Theory},''
\href{http://dx.doi.org/10.1016/j.physrep.2014.11.007}{{\em Phys. Rept.}
	{\bfseries 566} (2014) 1--60},
\href{http://arxiv.org/abs/1306.2643}{{\ttfamily arXiv:1306.2643 [hep-th]}}.

\bibitem{Aldazabal:2013sca}
G.~Aldazabal, D.~Marques, and C.~Nunez, 
``{Double Field Theory: A Pedagogical	Review},''
 \href{http://dx.doi.org/10.1088/0264-9381/30/16/163001}
 {{\em		Class. Quant. Grav.} 
	{\bfseries 30} (2013) 163001},
\href{http://arxiv.org/abs/1305.1907}
{{\ttfamily arXiv:1305.1907 [hep-th]}}.



\bibitem{Hohm:2013bwa}
O.~Hohm, D.~Lust, and B.~Zwiebach, ``{The Spacetime of Double Field Theory:
	Review, Remarks, and Outlook},''
\href{http://dx.doi.org/10.1002/prop.201300024}{{\em Fortsch. Phys.}
	{\bfseries 61} (2013) 926--966},
\href{http://arxiv.org/abs/1309.2977}{{\ttfamily arXiv:1309.2977 [hep-th]}}.
\bibitem{Park:2016sbw}
J.~H.~Park,
``Green-Schwarz superstring on doubled-yet-gauged spacetime,''
\href{http://dx.doi:10.1007/JHEP11(2016)005}{{\em JHEP} {\bfseries 11}
	(2016)005},
\href{http://arxiv.org/abs/1609.04265}{{\ttfamily arXiv:1609.04265 [hep-th]}}.



\bibitem{Hitchin:2003cxu}
N.~ Hitchin, ``Generalized Calabi-Yau manifolds,? 
\href{http://dx.doi:10.1093/qjmath/54.3.281}
{{\em Quart. J. Math. Oxford}	{\bfseries	 54} (2003)   281-308},
{{\ttfamily arXiv:math/020909  [math.DG]}}.

\bibitem{Hitchin:2005in}
N.~ Hitchin, ``Brackets, forms and invariant functionals,? 
{{\ttfamily arXiv:math/0508618 	 [math.DG]}}.

\bibitem{Gualtieri:2003dx}
M.~ Gualtieri, ''Generalized complex geometry,?
{{\ttfamily arXiv:math/0401221 	 [math.DG]}}.


\bibitem{Hull:1994ys}
C.~M.~Hull and P.~K.~Townsend,
``Unity of superstring dualities,''
\href{http://dx.doi:10.1016/0550-3213(94)00559-W}
{{\em Nucl. Phys.}	{\bfseries	 B438} (1995) 109-137},
\href{http://arxiv.org/abs/hep-th/9410167}
{{\ttfamily arXiv:hep-th/9410167		[hep-th]}}.

\bibitem{Witten:1995ex}
E.~Witten,
``String theory dynamics in various dimensions,''
\href{http://dx.doi:10.1016/0550-3213(95)00158-O}
{{\em Nucl. Phys.}	{\bfseries	 B443} (1995) 85-126},
\href{http://arxiv.org/abs/hep-th/9503124}
{{\ttfamily arXiv:hep-th/9503124		[hep-th]}}.


\bibitem{Vafa:1996xn}
C.~Vafa,
``Evidence for F theory,''
\href{http://dx.doi:10.1016/0550-3213(96)00172-1}
{{\em Nucl. Phys.}	{\bfseries	 B469} (1996)  403-418},
\href{http://arxiv.org/abs/hep-th/9602022}
{{\ttfamily arXiv:hep-th/9602022		[hep-th]}}.

\bibitem{Blencowe:1988sk}
M.~P.~Blencowe and M.~J.~Duff,
``Supermembranes and the Signature of Space-time,''
\href{http://dx.doi:10.1016/0550-3213(88)90155-1}
{{\em Nucl. Phys.}	{\bfseries	 B310} (1988)  387-404}.

\bibitem{Hull:1995xh}
C.~M.~Hull,
``String dynamics at strong coupling,''
\href{http://dx.doi:10.1016/0550-3213(96)00096-X}
{{\em Nucl. Phys.}	{\bfseries	 B468} (1996)  113-154 },
\href{http://arxiv.org/abs/hep-th/9512181}
{{\ttfamily arXiv:hep-th/9512181		[hep-th]}}.

\bibitem{Hull:2007zu}
C.~M.~Hull,
``Generalised Geometry for M-Theory,''
\href{http://dx.doi:10.1088/1126-6708/2007/07/079}{{\em JHEP} {\bfseries 07}
	(2007)079},
\href{http://arxiv.org/abs/hep-th/0701203}{{\ttfamily arXiv:hep-th/0701203 [hep-th]}}.

\bibitem{PiresPacheco:2008qik}
P.~Pires Pacheco and D.~Waldram,
``M-theory, exceptional generalised geometry and superpotentials,''
\href{http://dx.doi:10.1088/1126-6708/2008/09/123}{{\em JHEP} {\bfseries 09}
	(2008) 123},
\href{http://arxiv.org/abs/0804.1362}{{\ttfamily arXiv:0804.1362 [hep-th]}}.

\bibitem{Berman:2010is}
D.~S.~Berman and M.~J.~Perry,
``Generalized Geometry and M theory,''
\href{http://dx.doi:10.1007/JHEP06(2011)074}{{\em JHEP} {\bfseries 06}
	(2011) 074},
\href{http://arxiv.org/abs/1008.1763}{{\ttfamily arXiv:1008.1763 [hep-th]}}.

\bibitem{Berman:2011pe}
D.~S.~Berman, H.~Godazgar and M.~J.~Perry,
``SO(5,5) duality in M-theory and generalized geometry,''
\href{http://dx.doi:10.1016/j.physletb.2011.04.046}{{\em Phys. Lett.} {\bfseries B700}
	(2011) 65-67},
\href{http://arxiv.org/abs/1103.5733}{{\ttfamily arXiv:1103.5733 [hep-th]}}.

\bibitem{Hohm:2013vpa}
O.~Hohm and H.~Samtleben,
``Exceptional Field Theory I: $E_{6(6)}$ covariant Form of M-Theory and Type IIB,''
\href{http://dx.doi:10.1103/PhysRevD.89.066016}{{\em Phys. Rev.}
	{\bfseries	 D89} (2014)    066016 },
\href{http://arxiv.org/abs/1312.0614}
{{\ttfamily arXiv:1312.0614		[hep-th]}}.
%
\bibitem{Hohm:2013uia}
O.~Hohm and H.~Samtleben,
``Exceptional field theory. II. E$_{7(7)}$,''
\href{http://dx.doi:10.1103/PhysRevD.89.066017}{{\em Phys. Rev.}
	{\bfseries	 D89} (2014)    066017 },
\href{http://arxiv.org/abs/1312.4542}
{{\ttfamily arXiv:1312.4542		[hep-th]}}.
%
\bibitem{Hohm:2014fxa}
O.~Hohm and H.~Samtleben,
``Exceptional field theory. III. E$_{8(8)}$,''
\href{http://dx.doi:10.1103/PhysRevD.90.066002}{{\em Phys. Rev.}
	{\bfseries	 D90} (2014)   066002 },
\href{http://arxiv.org/abs/1406.3348}
{{\ttfamily arXiv:1406.3348		[hep-th]}}.
%

\bibitem{Coimbra:2011ky}
A.~Coimbra, C.~Strickland-Constable and D.~Waldram,
``$E_{d(d)} \times \mathbb{R}^+$ generalised geometry, connections and M theory,''
\href{http://dx.doi:10.1007/JHEP02(2014)054}{{\em JHEP} {\bfseries
		02} (2014) 054},
\href{http://arxiv.org/abs/1112.3989}{{\ttfamily arXiv:1112.3989 [hep-th]}}.

\bibitem{Berman:2012vc}
D.~S.~Berman, M.~Cederwall, A.~Kleinschmidt and D.~C.~Thompson,
``The gauge structure of generalised diffeomorphisms,''
\href{http://dx.doi:10.1007/JHEP01(2013)064}{{\em JHEP} {\bfseries
		01} (2013) 064},
\href{http://arxiv.org/abs/1208.5884}{{\ttfamily arXiv:1208.5884 [hep-th]}}.

\bibitem{Godazgar:2014nqa}
H.~Godazgar, M.~Godazgar, O.~Hohm, H.~Nicolai and H.~Samtleben,
``Supersymmetric E$_{7(7)}$ Exceptional Field Theory,''
\href{http://dx.doi:10.1007/JHEP09(2014)044}{{\em JHEP} {\bfseries
		09} (2014) 044},
\href{http://arxiv.org/abs/1406.3235}{{\ttfamily arXiv:1406.3235 [hep-th]}}.

\bibitem{Musaev:2015ces}
E.~T.~Musaev,
``Exceptional field theory: $SL(5)$,''
\href{http://dx.doi:10.1007/JHEP02(2016)012}{{\em JHEP} {\bfseries
		02} (2016) 012},
\href{http://arxiv.org/abs/1512.02163}{{\ttfamily arXiv:1512.02163 [hep-th]}}.

\bibitem{Abzalov:2015ega}
A.~Abzalov, I.~Bakhmatov and E.~T.~Musaev,
``Exceptional field theory: $SO(5,5)$,''
\href{http://dx.doi:10.1007/JHEP06(2015)088}{{\em JHEP} {\bfseries
		06}  (2015) 088},
\href{http://arxiv.org/abs/1504.01523}{{\ttfamily arXiv:1504.01523 [hep-th]}}.

\bibitem{Hatsuda:2012uk}
M.~Hatsuda and T.~Kimura,
``Canonical approach to Courant brackets for D-branes,''
 \href{http://dx.doi:10.1007/JHEP06(2012)034}{{\em JHEP}
	{\bfseries 06} (2012) 034},
\href{http://arxiv.org/abs/1203.5499}{{\ttfamily arXiv:1203.5499 [hep-th]}}.

\bibitem{Hatsuda:2012vm}
M.~Hatsuda and K.~Kamimura,
``SL(5) duality from canonical M2-brane,''
 \href{http://dx.doi:10.1007/JHEP11(2012)001}{{\em JHEP}
	{\bfseries 11} (2012) 001},
\href{http://arxiv.org/abs/1208.1232}{{\ttfamily arXiv:1208.1232 [hep-th]}}.

\bibitem{Hatsuda:2013dya}
M.~Hatsuda and K.~Kamimura,
``M5 algebra and SO(5,5) duality,''
 \href{http://dx.doi:10.1007/JHEP06(2013)095}{{\em JHEP}
	{\bfseries 06} (2013) 095},
\href{http://arxiv.org/abs/1305.2258}{{\ttfamily arXiv:1305.2258 [hep-th]}}.

\bibitem{Linch:2016ipx}
W.~D.~Linch and W.~Siegel,
``F-brane Dynamics,''
\href{http://arxiv.org/abs/1610.01620}{{\ttfamily arXiv:1610.01620 [hep-th]}}.

\bibitem{Ju:2016hla}
C.~Y.~Ju and W.~Siegel,
``Gauging Unbroken Symmetries in F-theory,''
\href{http://dx.doi:10.1103/PhysRevD.94.106004}{{\em Phys. Rev.}
	{\bfseries	 D94} (2016)  106004 },
\href{http://arxiv.org/abs/1607.03017}
{{\ttfamily arXiv:1607.03017		[hep-th]}}.

\bibitem{Linch:2015fca}
W.~D.~Linch and W.~Siegel,
``Critical Super F-theories,''
\href{http://arxiv.org/abs/1507.01669}{{\ttfamily arXiv:1507.01669 [hep-th]}}.

\bibitem{Linch:2015qva}
W.~D.~Linch and W.~Siegel,
``F-theory with Worldvolume Sectioning,''
 \href{http://dx.doi:10.1007/JHEP04(2021)022}{{\em JHEP}
	{\bfseries 04}(2021) 022},
\href{http://arxiv.org/abs/1503.00940}{{\ttfamily arXiv:1503.00940 [hep-th]}}.

\bibitem{Linch:2015fya}
W.~D.~Linch, III and W.~Siegel,
``F-theory from Fundamental Five-branes,''
 \href{http://dx.doi:10.1007/JHEP02(2021)047}{{\em JHEP}
	{\bfseries 02}(2021) 047},
\href{http://arxiv.org/abs/1502.00510}{{\ttfamily arXiv:1502.00510 [hep-th]}}.

\bibitem{Linch:2015lwa}
W.~D.~Linch and W.~Siegel,
``F-theory Superspace,''
\href{http://arxiv.org/abs/1501.02761}{{\ttfamily arXiv:1501.02761 [hep-th]}}.

\bibitem{Siegel:2016dek}
W.~Siegel,
``F-theory with zeroth-quantized ghosts,''
\href{http://arxiv.org/abs/1601.03953}{{\ttfamily arXiv:1601.03953 [hep-th]}}.

\bibitem{Siegel:2019wrr}
W.~Siegel and D.~Wang,
``F-theory superspace backgrounds,''
\href{http://arxiv.org/abs/1910.01710}{{\ttfamily arXiv:1910.01710 [hep-th]}}.

\bibitem{Siegel:2018puf}
W.~Siegel and D.~Wang,
``Enlarged exceptional symmetries of first-quantized F-theory,''
\href{http://arxiv.org/abs/1806.02423}{{\ttfamily arXiv:1806.02423 [hep-th]}}.

\bibitem{Siegel:2020qef}
W.~Siegel and D.~Wang,
``M Theory from F Theory,''
\href{http://arxiv.org/abs/2010.09564}{{\ttfamily arXiv:2010.09564 [hep-th]}}.

\bibitem{Pasti:1997gx}
P.~Pasti, D.~P.~Sorokin and M.~Tonin,
``Covariant action for a D = 11 five-brane with the chiral field,''
\href{http://dx.doi:10.1016/S0370-2693(97)00188-3}{{\em Phys. Lett.}
	{\bfseries	 B398} (1997)  41-46 },
\href{http://arxiv.org/abs/hep-th/9701037}
{{\ttfamily arXiv:hep-th/9701037		[hep-th]}}.

\bibitem{Siegel:1983es}
W.~Siegel, ``{Manifest Lorentz Invariance Sometimes Requires Nonlinearity},''
\href{http://dx.doi.org/10.1016/0550-3213(84)90453-X}{{\em Nucl. Phys.}
  {\bfseries B238} (1984) 307--316}.


\bibitem{Hatsuda:2020buq}
M.~Hatsuda, S.~Sasaki and M.~Yata,
``Five-brane current algebras in type II string theories,''
\href{http://doi:10.1007/JHEP03(2021)298}{{\em JHEP}
	{\bfseries 03} (2021) 298},
\href{http://arxiv.org/abs/2011.13145}{{\ttfamily arXiv:2011.13145 [hep-th]}}.



\end{thebibliography}
\end{document}